\documentclass{iopart}

\expandafter\let\csname equation*\endcsname\relax 
\expandafter\let\csname endequation*\endcsname\relax
\usepackage{amsmath}
\usepackage{amsfonts}
\usepackage{amssymb}
\usepackage{amsthm}

\usepackage[noprefix,intoc,refpage]{nomencl}
\usepackage{pst-plot}
\usepackage{ltxtable}
\usepackage{textcomp}
\usepackage{tabularx}
\usepackage{graphicx, color}
\usepackage{subfigure}

\usepackage{fixmath}
\usepackage{tensind}
\usepackage{wasysym}
\usepackage{afterpage}
\usepackage{dsfont}
\usepackage{bibgerm}
\usepackage{hyperref}
\usepackage{slashed}
\usepackage{mathrsfs}
\usepackage{array}
\usepackage{makeidx}
\usepackage{enumerate}
\makeindex

\def\e{\mathrm{e}}
\def\ii{\mathrm{i}}

\def\p{\partial}
\def\bea{\begin{eqnarray}}
\def\eea{\end{eqnarray}}
\def\qbin#1#2{\genfrac{[}{]}{0pt}{}{#1}{#2}}

\def\psg{\phantom{-}}
\def\nequiv{\not\equiv}
\def\trans{\!{\!{t}}}
\def\Hi{{\cal H}}
\def\Cop{{\mathbb C}}
\def\Zop{{\mathbb Z}}

\def\Nop{{\mathbb N}}
\def\Rc{{\cal R}} 

\def\S{\mathfrak{S}}

\whenindex{w}{\mathbf{a}}{}
\whenindex{x}{\mathbf{b}}{}
\whenindex{y}{\mathbf{c}}{}
\whenindex{z}{\mathbf{d}}{}
\whenindex{W}{\mathbf{A}}{}
\whenindex{X}{\mathbf{B}}{}
\whenindex{Y}{\mathbf{C}}{}
\whenindex{Z}{\mathbf{D}}{}
\whenindex{'}{'}{}

\newcommand{\ket}[1]{|#1\rangle}

\def\bsm{\left( \!\begin{smallmatrix}}
\def\esm{\end{smallmatrix} \!\right)}

\DeclareMathOperator{\rog}{L}

\renewcommand{\sp}[1]{\left\langle #1 \right\rangle}

\tensordelimiter{?}

\numberwithin{equation}{section}

\graphicspath{{graphics/}}

\newcommand{\Th}{\Theta}
\newcommand{\eps}{\epsilon}

\newcommand{\Z}{\mathbb{Z}}

\renewcommand{\P}{\mathbb{P}}

\newcommand{\C}{\mathbb{C}}
\newcommand{\Q}{\mathbb{Q}}
\newcommand{\N}{\mathbb{N}}
\renewcommand{\l}{\left(}
\renewcommand{\r}{\right)}

\newcommand{\2}{\frac{1}{2}}

\newcommand{\be}{\begin{equation}}
\newcommand{\ee}{\end{equation}}

\renewcommand{\pmod}[1]{\ (\mathrm{mod} \ #1)}
\newcommand{\vin}[1]{\in (\Z_{\geq 0})^{#1}}
\newcommand{\mac}[1]{\mathcal{#1}}
\def\nn{\notag}

\begin{document}\allowdisplaybreaks
\selectlanguage{english}

\review[]{\mbox{What the characters of irreducible subrepresentations}\\ of Jordan cells can tell us about LCFT}
\author{Michael Flohr$^1$ and Michael Koehn$^2$}
\address{$^1$ Institute for Theoretical Physics, Leibniz University Hannover, Appelstra\ss e 2, 30167 Hannover, Germany, EU
}
\address{$^2$ Max-Planck-Institut f\"ur Gravitationsphysik, Albert-Einstein-Institut, 
Am M\"uhlenberg 1, 14476 Potsdam, Germany, EU
}
\ead{michael.flohr@itp.uni-hannover.de, michael.koehn@aei.mpg.de}

\begin{abstract}
In this article, we review some aspects of logarithmic conformal
field theories which can be inferred from the characters of irreducible 
submodules of indecomposable modules. We will mainly consider the
$\mathcal{W}(2,2p-1,2p-1,2p-1)$ series of triplet algebras and a bit
logarithmic extensions of the minimal Virasoro models. Since in all known
examples of logarithmic conformal field theories the vacuum representation 
of the maximally extended chiral symmetry algebra is
an irreducible submodule of a larger, indecomposable module, its character
provides a lot of non-trivial information about the theory such as a set of
functions which spans the space of all torus amplitudes. Despite such
characters being modular forms of inhomogeneous weight, they fit in the
$ADET$-classification of fermionic sum representations. Thus, they show that
logarithmic conformal field theories naturally have to be taken into account
when attempting to classify rational conformal field theories.
\end{abstract}

\submitto{J.Phys.A: Special Issue on ``Logarithmic Conformal Field Theory''}
\maketitle

\pagenumbering{arabic} \setcounter{page}{1}

\section{Introduction}

Suppose you have constructed a ${\cal W}$-algebra, i.e.\ a maximally extended
symmetry algebra of a two-dimensional conformal field theory. Suppose you 
study its vacuum sector and count the states on each of the first $N$ levels.
Thus you compute the first $N$ orders of the character of its vacuum
representation. Assuming $N$ large enough, what can you infer from this?

The point is that the ${\cal W}$-algebra and its vacuum sector are purely
algebraic objects which can, in principle, computed with computer algebra
systems. But a lot more of the conformal field theory is already encoded in 
these structures which, by comparison, are relatively easy to come by. 

If we have reason to believe that the conformal field theory under consideration
is rational, knowledge of the vacuum character to a sufficient high order $N$ 
is enough to find all other admissible Virasoro highest weights and,
furthermore, a set of functions which span the space of the vacuum torus 
amplitudes of the theory. In particular, you can decide from the vacuum
character whether the theory is logarithmic or not. The reason behind this is,
that the vacuum torus amplitudes of a rational (logarithmic) conformal field
theory form a representation of the modular group, and that a modular form
is already uniquely determined by a certain, finite, number of terms of its
$q$-expansion. 

The characters of irreducible modules, which are submodules of larger,
indecomposable modules, do tell us even more. Since these characters turn out
to be modular forms which consist of parts with different modular weights, i.e.
they are not modular forms of homogeneous modular weight, their modular
transforms give rise to functions, which are not characters of other
representations, but can be understood as torus vacuum amplitudes. 

It has been conjectured that the characters of rational
conformal field theories, which are modular forms, all have fermionic sum
representations which can all be found within the so-called 
$ADET$-classification of fermionic sums (cf.~\cite{Nah04}). As it turns out, this holds true even
for the characters of such irreducible submodules of indecomposable modules. 
And that means that rational logarithmic conformal field theories seem to
naturally fit into these classification schemes of rational conformal field 
theories. 

Our considerations are mainly based on the relatively well understood
$\mathcal{W}(2,2p-1,2p-1,2p-1)$ series of triplet algebras. However, as long
as the vacuum representation is an irreducible submodule of an indecomposable
module, our reasoning and strategies will continue to work. In particular, we
expect that augmented minimal models can be explored along the lines set out
in our paper. Especially, the space of torus vacuum amplitudes can be 
determined from which, with some further assumptions or some more data from
explicitly constructed representations, the characters and fusion rules could
be derived. 

\subsubsection*{Bosonic-Fermionic $q$-Series Identities}

The zero mode $L_0$ of the infinite-dimensional Virasoro symmetry algebra of
two-dimensional conformal quantum field theory admits a natural gradation of
the representation space into subspaces of fixed $L_0$ eigenvalue. The
character of a representation displays the number of linearly independent
states in each of these subspaces as a series expansion in some formal variable
$q$, where the dimension $k$ of the subspace with fixed $L_0$ eigenvalue $n$ is
indicated by a summand $kq^n$ in this series. Non-unique realizations of the
state spaces in two-dimensional conformal field theories imply the existence of
several alternative character formulae. The Bose-Fermi correspondence indicates
that the characters of two-dimensional quantum field theories can be expressed
in a \emph{bosonic} as well as in a \emph{fermionic} way, leading to
\emph{bosonic-fermionic $q$-series identities}\index{bosonic-fermionic
$q$-series identity}. The existence of such identities can be traced back to
the famous \emph{Rogers-Ramanujan identities}\index{Rogers-Schur-Ramanujan
identities} \cite{Rog94,Sch17,RR19}
\be
\label{rsr-intro}
\sum\limits_{n=0}^{\infty}\frac{q^{n(n+a)}}{(q)_n} = \prod\limits^{\infty}_{n=1}
\frac{1}{(1-q^{5n-1-a})(1-q^{5n-4+a})}
\ee
for $ a\in\{0,1\} $. For the notations, see section three, 
equations \eqref{rsr} ff.
These identities later turned out to coincide with the two characters of the
$\mathcal{M}(5,2)$\nomenclature[Mpp]{$\mathcal{M}(p,p')$}{The minimal 
Virasoro model with central charge $c_{p,p'}$, $p,p'\in\Z_{\geq 2}$ coprime} 
minimal model \cite{franco} (up to an overall factor $q^{\alpha}$ for 
some $\alpha\in\Q$). By using Jacobi's triple product identity 
(see e.g.~\cite{And84}), the right hand side of \eqref{rsr-intro} 
can be transformed to give a simple example of a 
\emph{bosonic-fermionic $q$-series identity}:
\be \label{bf_id-intro}
\sum\limits_{n=0}^{\infty}\frac{q^{n(n+a)}}{(q)_n} =  \frac{1}{(q)_{\infty}}\sum\limits_{n=-\infty}^{\infty}(q^{n(10n+1+2a)}-q^{(5n+2-a)(2n+1)}) \ .
\ee
The bosonic expression on the right hand side of \eqref{bf_id-intro} corresponds to two special cases of the general character formula for minimal models $\mathcal{M}(p,p')$ \cite{RoC84} 
\be
\label{generalcharacter}
\hat{\chi}^{p,p'}_{r,s}=q^{\frac{c}{24}-h^{p,p'}_{r,s}}\chi^{p,p'}_{r,s}=\frac{1}{(q)_{\infty}}\sum\limits_{n=-\infty}^{\infty}(q^{n(npp'+pr-p's)}-q^{(np+s)(np'+r)})
\ee 
with $\hat{\chi}^{p,p'}_{r,s}$\nomenclature[chipprs]{$\chi_{r,s}^{p,p'}$}{The character of the minimal Virasoro model $\mathcal{M}(p,p')$ corresponding to Kac indices $(r,s)$} being the normalized character.
It has been termed bosonic in \cite{KKMM93b} because it is computed by eliminating null states from a Verma module over the Virasoro algebra.
The signature of bosonic character expressions is the alternating sign, which reflects the subtraction of null vectors, whereas each summand of a fermionic character formula is manifestly positive. 
Furthermore, the factor $(q)_\infty$ keeps track of the free action of the Virasoro ``creation'' modes.

\subsubsection*{Quasi-Particle Interpretation for Fermionic Character Expressions}

On the other hand, the fermionic sum representation for a character possesses a remarkable interpretation in terms of an underlying system of \emph{quasi-particles}. These expressions first occurred on the left hand side of the Rogers-Ramanujan identities \eqref{rsr-intro}. Generalizations have been obtained by Andrews and Gordon \cite{And74b,Gor61} and later on by Lepowsky and Primc \cite{LP85}. The most general fermionic expression is regarded to be a linear combination of fundamental fermionic forms.
A \emph{fundamental fermionic form} \cite{BMS98,Wel05,DKMM94} is\footnote{The constant $c$ is not to be confused with the central charge $c_{p,p'}$.}
\be
\label{fff}\sum_{\substack{\vec{m}\in(\mathbb{Z}_{\geq 0})^r \\ \text{restrictions}}}\frac{q^{\vec{m}^tA\vec{m}+\vec{b}^t\vec{m}+c}}{\prod_{i=1}^{j}(q)_i}\prod_{i=j+1}^{r}\qbin{g(\vec{m})}{m_i}_q
\ee
with $A\in M_r(\mathbb{Q})$, $\vec{b}\in\mathbb{Q}^r$, $c\in\mathbb{Q}$, $0 \leq j \leq r$, $g$ a certain linear, algebraic function in the $m_i, 1\leq i \leq r$, and the $q$-deformed binomial coefficient (the \emph{$q$-binomial coefficient}) defined as 
\be
\begin{bmatrix} n \\ m \end{bmatrix}_q=\begin{cases} \frac{(q)_{n}}{(q)_{m}(q)_{n-m}} & \mbox{if} \; \; 0 \le m \le n \; \; \\ 0 & \mbox{otherwise} \; \; \\ \end{cases} .
\ee
The sum over $\vec m$ is an abbreviation and implies that each component $m_i$ 
of $\vec m$ is to be summed over independently.

It is sometimes also called \emph{universal chiral partition function}\index{universal chiral partition function} for exclusion statistics \cite{BM98,Sch99}. Remarkably, in most cases the matrix $A$ is given by the \emph{Cartan matrix} of a simple Lie algebra or by its inverse. It turns out that \eqref{fff} can be interpreted in terms of a system of $r$ different species of fermionic quasi-particles with non-trivial momentum restrictions. The bosonic representations are in general unique, whereas there is usually more than one fermionic expression for the same character, giving rise to different quasi-particle interpretations for the same conformal field theory which are conjectured to correspond to different integrable massive extensions of the theory, see section \ref{ref:adet}. There are cases for which such correspondences are known. Thus, the different interpretations in terms of quasi-particle systems may be a guide for experimental research. Note that in general, the existence of quasi-particles has 
been experimentally demonstrated, namely in the case of the fractional quantum Hall effect \cite{SGJE97}. They turned out to be of charge $e/3$, as predicted by Laughlin \cite{Lau83,TSG82}.

\subsubsection*{Dilogarithm Identities}\label{ref:dilog}

Furthermore, knowledge of a fermionic character expression and either the theory's effective central charge or a product form of the character results in dilogarithm identities (see also \cite{Nah04,Zag07,Nak11}) of the form
\be
\frac{1}{\rog(1)}\sum_{i=1}^{N}\rog(x_i)=d \ ,
\ee
where $\rog$ is the Rogers dilogarithm \cite{Rog07,Lew58,Lew81b} and $x_i$ and $d$ are rational numbers. It is conjectured \cite{NRT93} that all values of the effective central charges occurring in non-trivial rational conformal field theories can be expressed as one of those rational numbers that consist of a sum of an arbitrary number of dilogarithm functions evaluated at algebraic numbers from the interval $(0,1)$. Besides their intriguing relation to wall-crossing formulae in string theory \cite{APP11}, dilogarithm identities are well-known to arise from thermodynamic Bethe ansatz. Conversely, there is also a conjecture \cite{Ter92} that dilogarithm identities corresponding to Bethe ansatz equations $x_i=\prod_{j=1}^k (1-x_j)^{2A_{ij}}$, where $A$ is the inverse Cartan matrix of one of the $ADET$ series of simple Lie algebras (see section \ref{ref:adet}), imply fermionic character expressions of rational conformal field theory characters. Thus, the study of dilogarithm identities arising from conformal 
field theories gives further insight into the classification of all rational theories.

\subsubsection*{Fermionic Expressions as Evidence for Rationality of Extended $\mathcal{W}$-Symmetry Theories}

In addition to the minimal models of the Virasoro algebra, there exist other theories endowed with more symmetries. They are generated by modes of currents different from the energy-momentum tensor. Possible extensions, which contain the Virasoro algebra as a subalgebra, lead to free fermions, Kac-Moody algebras \cite{kac}, Superconformal algebras \cite{GKO86} or more generally $\mathcal{W}$-algebras\nomenclature[W]{$\mathcal{W}$}{An algebra which includes the Virasoro algebra and may be extended by the modes of additional fields} \cite{Nah89,BS95}. Amongst others, the $\mathcal{W}(2,2p-1,2p-1,2p-1)$ series of conformal field theories with extended triplet algebra symmetry \cite{Kau91} is investigated in this report, comprising the best understood examples of \emph{logarithmic} conformal field theory models. These models correspond to central charges $c_{p,1}=1-6\frac{(p-1)^2}{p}$, $p\geq 2$. In fact, the description of various physical processes in two dimensions requires logarithmic divergencies in the 
correlation functions, as in the case of e.g.~magnetohydrodynamics \cite{ST98}, turbulence \cite{RR96}, dense polymers \cite{PR06} or percolation \cite{RP07,FM05}. For other interesting applications, also see \cite{FLN06,FLN07} for a work concerning instantons, where a non-diagonalizable Hamiltonian may occur, and also the recent results of \cite{BdHMRZ12} about three-dimensional tricritical gravity.

For some values of the central charge (when there are fields with integer-spaced dimensions), the existence of fields that lead to logarithmic divergencies in four-point functions is unavoidable \cite{Gur93}. Conventionally, if the state space of states of a conformal field theory decomposes into a finite sum of irreducible representations, then the theory is said to be \emph{rational}\index{rational conformal field theory}. However, the investigation of certain classes of logarithmic conformal field theories pointed towards loosening this strict definition of rationality to also include reducible, but indecomposable representations. Recently, an attempt at organizing logarithmic theories into families alongside related rational theories has been started \cite{EF06b,FGST06c,PRZ06}. For logarithmic conformal field theories, almost all of the basic notions and tools of (rational) conformal field theories, such as null vectors, (bosonic) character functions, partition functions, fusion rules, modular invariance,
 have been generalized by now \cite{Flo03}, the main difference to ordinary rational conformal field theories such as the minimal models remaining the occurrence of indecomposable representations.

In this light, the existence of a complete set of fermionic sum representations for the characters of the $\mathcal{W}(2,2p-1,2p-1,2p-1)$ logarithmic conformal field theory models with $p\geq 2$ (which are also referred to as the $c_{p,1}$\nomenclature[cpp]{$c_{p,p'}$}{The central charge $c_{p,p'}=1-6\frac{(p-p')^2}{pp'}$ of the minimal Virasoro model $\mathcal{M}(p,p')$} models) provides further evidence that these models, although outside of the usual classification scheme of rational conformal field theories, are nonetheless bona fide theories \cite{FGK07}.


\section{Character Expressions for Conformal Field Theories with $\mac{W}$-symmetry}

\subsection{Virasoro Characters} \label{ref:chars}

Throughout this review, we are only interested in one chiral half of the CFT.
Thus, we consider representations $\mathcal{H}(h,c)$ of the chiral symmetry
algebra, which may either be irreducible representations 
(often denoted ${\cal V}_h$ or ${\cal W}(h)$), or 
indecomposable ones (for example denoted ${\cal R}_h$ or 
${\cal R}(h,h_1\ldots)$). We stress that, if there are indecomposable 
representations involved, the state space will not factorize into a direct sum
of tensor products of holomorphic and corresponding anti-homomorphic sectors.

The \emph{character}\index{character} of an indecomposable $\mathrm{Vir}$-module $V$ of the Virasoro algebra $\mathrm{Vir}$ is defined by
\be \label{char-general}
\chi_V(\tau):=\Tr\e^{2\pi\ii\tau(L_0-\frac{c}{24})} \ .
\ee
with $c$ being the central charge and $L_0$ the Virasoro zero mode. 
The characters of the representations are an essential ingredient for a
conformal field theory. Since $L_0$ corresponds to the Hamiltonian of the
(chiral half of the) system, the energy spectrum (at least certain sectors) is
encoded in the character. The trace is usually taken over an irreducible
highest weight representation and the factor $q^{-\frac{c}{24}}$ guarantees the
needed linear behavior under modular transformations.
By setting $q:=\e^{2\pi\ii\tau}$, \eqref{char-general} leads to
\be
\chi_V(q)=q^{-\frac{c}{24}}\sum_hq^h\text{dim eigenspace}(L_0^d,h) \ ,
\ee
where $L_0^d$ is the diagonalizable summand of the possibly non-diagonalizable
$L_0$. To compute the character of a Verma module $V(h,c)$, we have to compute the
number of linearly independent states at a given level $k$. We can grade the
Verma module by its $L_0$ eigenvalue:
\be
V(h,c)=\bigoplus_N V_N(h,c)
\ee
with
\be
V_N(h,c)=\sp{\left\{L_{-n_1}L_{-n_2}\cdots L_{-n_k}\ket{h}\mid k\in\Z_{\geq 0},\ n_{i+1} \geq n_i,\ \sideset{}{_{i=1}^k}\sum n_i=N\right\}}\ .
\ee
Thus, the number of distinct, linear independent states at level $N$ is given
by the number $p(N)$ of additive partitions of the integer $N$. The generating
function for the number of partitions is
\be
\frac{1}{\phi(q)}\equiv\prod_{n=1}^{\infty}\frac{1}{1-q^n}=\sum_{n=0}^{\infty}p(N)q^n \ ,
\ee
where $\phi(q)$ is the Euler function. Hence the character is given by
\be
\chi_{V(h,c)}=q^{h-\frac{c}{24}}\sum_{k=0}^{\infty}p(k)q^k \ .
\ee
Dedekind's $\eta$ function $\eta(q)=q^{1/24}\phi (q)$ is conventionally used
because it simplifies the analysis of a character function under modular
transformations:
\be
\chi_{V(h,c)}=\frac{q^{\frac{1-c}{24}}}{\eta(q)}q^h \ .
\ee
This series is convergent if $|q|<1$, i.e.~$\tau\in\mathbb{H}$ (upper
half-plane). If $V(h,c)$ already is an irreducible representation, i.e.~is
non-degenerate, then this is its character. If not, the characters $\chi_{r,s}$
of irreducible representations $M(h_{r,s},c)$ can be read off the corresponding
embedding structure \cite{FF83,Flo93}.

\subsection{$\Theta$- and $\eta$-functions and their Modular Transformation Properties} \label{ref:theta}

The \emph{Jacobi-Riemann $\Theta$-functions}\index{Jacobi-Riemann
$\Theta$-functions} and the \emph{affine $\Theta$-functions}\index{affine
$\Theta$-functions} are defined by
\begin{align}
\label{theta}\Theta_{\lambda,k}(\tau) & =\sum\limits_{n\in \mathbb Z}q^{\frac{(2kn+\lambda)^2}{4k}}
\intertext{and}
\label{deltheta}(\partial \Theta)_{\lambda,k}(\tau) & =\sum\limits_{n\in \mathbb Z}(2kn+\lambda)q^{\frac{(2kn+\lambda)^2}{4k}}
\end{align}
with $q=\e^{2\pi\ii\tau}$. $\lambda\in\frac{\Z}{2}$ is called the \emph{index}
and $k\in\frac{\Z_{\geq 1}}{2}$ the \emph{modulus}. The $\Theta$-functions
satisfy the symmetries 
\begin{align}
\label{theta_symmetries}\Theta_{\lambda,k}=\Theta_{-\lambda,k}=\Theta_{\lambda+2k,k} \quad\text{and} \\
\label{deltheta_symmetries}(\partial\Theta)_{-\lambda,k}=-(\partial\Theta)_{\lambda,k} \ .
\end{align}
The \emph{Dedekind $\eta$-function}\index{Dedekind $\eta$-function} is defined
as
\be\label{etadef}
\eta(\tau) = q^{\frac{1}{24}}\prod_{n=1}^{\infty}(1-q^n) \ .
\ee
The Jacobi-Riemann $\Theta$-functions and the Dedekind $\eta$-function are
modular forms of weight $1/2$, while the affine $\Theta$-functions have modular
weight $\frac{3}{2}$.
A \emph{modular form of weight $k$}\index{modular form} is defined by the
relation
\be
f\left( \frac{a\tau +b}{c\tau + d}\right)=\eps (a,b,c,d ) (c\tau +d)^kf(\tau)
\ee
for $\tau\in\C$ and $|\eps (a,b,c,d)|=1$ and with $\bsm a & b \\ c & d \esm \in
\Gamma\equiv \mathrm{PSL}(2,\Zop)$ and $f$ being a holomorphic function on the upper half-plane which is
also holomorphic at the \emph{cusp}\index{cusp}, i.e.~is holomorphic as
$\tau\rightarrow \ii \infty$.
The modular transformation properties of the $\Theta$- and $\eta$-functions for
those cases of $\lambda$ and $k$ that are needed in this report are
\begin{align}
\Theta_{\lambda,k}(-\frac{1}{\tau}) & = \sqrt{\frac{-\ii\tau}{2k}}\sum_{\lambda'=0}^{2k-1}\e^{\ii\pi\frac{\lambda\lambda'}{k}}\Theta_{\lambda',k} (\tau) \quad\text{for }\lambda\in\Z \\
\Theta_{\lambda,k}(\tau +1) & = \e^{\ii\pi\frac{\lambda^2}{2k}}\Theta_{\lambda,k} (\tau) \quad\text{for }\lambda -k\in\Z \\
\eta(-\frac{1}{\tau}) & = \sqrt{-\ii\tau}\eta(\tau) \\
\eta(\tau +1) & = \e^{\frac{\pi\ii}{12}}\eta(\tau) \ .
\end{align}
The functions $\chi_{\lambda,k}=\frac{\Theta_{\lambda,k}}{\eta}$, which often
turn up as summands in character functions in chapter \ref{ref:chars}, are
thus modular forms of weight zero with respect to the principal congruence subgroup
$\Gamma (N)$ of the modular group $\mathrm{PSL}(2,\Zop)$. The principal congruence subgroup is defined such that the diagonal matrix entries are congruent to 1 mod N, and the off-diagonal entries to 0 mod N. Many details about
$\Theta$-functions may be found in \cite{Igu72,Akh90}.


\subsection{Torus amplitudes} \label{ref:torus}

In a seminal work \cite{Zhu96}, Zhu proved several general facts about
conformal field theories or vertex operator algebras, whose characters form
finite dimensional representations of the modular group. Of particular interest
are theories, which satisfy the $C_2$ condition, but we shall not assume that 
they define {\it rational} conformal field theories. 
As it is somehow beyond the scope of this review to discuss, how a
mathematical rigorous definition of rationality of a CFT should incorporate 
the case of logarithmic conformal field theories, it will be sufficient, to 
appeal to the standard CFT lore of rationality. Thus, for the purpose of this
review, we call a conformal field theory rational if its chiral symmetry 
algebra possesses finitely many irreducible representations,
each of which has finite-dimensional $L_0$ eigenspaces, such that their fusion
products decompose into direct sums of just these irreducible representations.
The $C_2$ condition states
that the quotient space $\Hi_0 / C_2(\Hi_0)$ is finite dimensional,
where $\Hi_0$ is the vacuum representation of the conformal field
theory and $C_2(\Hi_0)$ is the space spanned by the states 
\be\label{span}
V_{-h(\psi)-1}(\psi)\, \chi \,, \qquad \hbox{for $\psi,\chi\in\Hi_0$.}
\ee
The $C_2$ condition implies that Zhu's algebra $A(\Hi_0)$ is finite
dimensional, and therefore that the conformal field theory has only
finitely many irreducible highest weight representations (see 
\cite{GG00} for an introduction to these matters and what is meant by
highest weight representations in case of extended chiral symmetry algebras). 
However, it does
not imply that the theory is rational, neither in the above sense, nor in the 
sense given by a mathematical rigorous definition of rationality. Indeed, the
prime examples of logarithmic theories, the triplet algebras at
$c=c_{p,1}$ \cite{Kau91}, do satisfy the $C_2$ condition \cite{CF06},
yet are not rational since they possess indecomposable
representations \cite{GK96b}.

As shown by Zhu,
if the conformal field theory satisfies the $C_2$ condition, then
every highest weight representation for which $L_0$ is diagonalizable
with finite dimensional eigenspaces gives rise to a torus amplitude.
In particular, the vacuum amplitude is just given by the usual character 
\be\label{char}
\chi_{\Hi_j}(\tau) = 
\hbox{Tr}_{\Hi_j} \left( q^{L_0-{c\over 24}} \right) \,, \qquad
q= e^{2\pi i \tau} \,,
\ee
which converges absolutely for $0 < |q| < 1$. Again, if the $C_2$
condition is satisfied, the space of torus amplitudes is finite
dimensional, and it carries a representation of
$\mathrm{PSL}(2,\Zop)$.  

Furthermore, Zhu points out that, if the
conformal field theory satisfies the $C_2$ condition, then there
exists a positive integer $s$ so that for a certain redefined algebra 
\be\label{condition}
L[-2]^s \, \Omega 
+ \sum_{r=0}^{s-1} g_r(q) \, L[-2]^r\, \Omega \in O_q(\Hi_0) 
\,,
\ee
where the modes $L[-2]$ are defined in equation (4.2.3) of Zhu's work 
\cite{Zhu96}, proof of his theorem 4.4.1, 
and $O_q(\Hi_0)$ denotes the subspace
of $\Hi_0$ whose one-point torus functions vanish. Here $g_r(q)$ are
polynomials in the Eisenstein series $E_4(q)$ and $E_6(q)$; we
shall choose the convention that the Eisenstein series are defined by 
\begin{align}
E_k(q) &= 1 - \frac{2\, k}{B_k}
\sum_{n=1}^{\infty}\sigma_{k-1}(n)\, q^n\,,\\
\sigma_k(n) &= \sum_{d|n}d^k\,,
\end{align}
where $B_k$ is the $k$-th Bernoulli number. Thus, the $q$-expansion of
the Eisenstein series reads $E_2 = 1 - 24q-72q^2-96q^3-\cdots$,  
$E_4 = 1 + 240q + 2160q^2 + 6720q^3+\cdots$, and 
$E_6 = 1 - 504q - 16632q^2 -122976q^3-\cdots$ in our normalization.

Given Zhu's definition of  $\tilde{\omega} \ast_{\tau} \psi$ (see equation (5.3.1) in \cite{Zhu96}, p.292, together with $L[n]=\tilde{\omega}[n+1]$), 
one can rewrite (\ref{condition}) as  
\be\label{condition1}
\tilde{\omega}^s \ast_{\tau} \Omega + 
\sum_{r=0}^{s-1} h_r(q)\, \tilde{\omega}^r \ast_{\tau} \Omega
\in O_q(\Hi_0) \,,
\ee
where now $h_r(q)$ are polynomials in the Eisenstein series
$E_2(q)$, $E_4(q)$ and $E_6(q)$, and we have used the notation 
\be\label{prod}
\tilde{\omega}^r \ast_{\tau} \Omega \equiv
\underbrace{\tilde{\omega} \ast_{\tau} \cdots
\ast_{\tau} \tilde{\omega} 
}_{r}\ast_{\tau}\Omega \,.
\ee
This has a few important consequences. 
Firstly, it therefore follows that every torus vacuum
amplitude $T(q)$ must satisfy the differential equation   
\be\label{diff}
\left[ \left(q {{\rm d}\over {\rm d}q}\right)^s + \sum_{r=0}^{s-1} h_r(q) 
 \left(q {{\rm d}\over {\rm d}q}\right)^r \right] T(q) = 0 \,.
\ee
Secondly, it furthermore follows that the functions $h_r$ are such that 
\be\label{rel}
\left(L[0]-{c\over 24}\right)^s + \sum_{r=0}^{s-1} h_r(0) 
\left(L[0]-{c\over 24}\right)^r = 0 
\ee
in Zhu's algebra $A(\Hi_0)$ (that is defined in section~2 of
\cite{Zhu96}). As we shall argue later, the differential equation
(\ref{diff}) can be identified with the modular differential equation
that was first considered in \cite{MMS1,MMS2}.  

If the conformal field theory is in addition rational in the above
sense Zhu showed that the space of torus amplitudes is spanned
by the characters of the irreducible representations,
and therefore that their characters transform into one another
under the action of the modular group. In case of logarithmic triplet 
neither of these two statements is correct. In fact, the space of 
torus vacuum amplitudes turns out to be larger than the space spanned by the
characters. Only the former forms a closed representation of the modular group.
Within the latter, there exists a smaller subset of characters, which forms a
smaller representation of the modular group, if one omits all characters of
submodules of indecomposable modules. However, restricting to this smaller
set of characters precisely loses all information about the theory that would
show that it is logarithmic. It is plausible that these considerations, which 
are explicitly proven for the triplet algebras, are generally valid for 
$C_2$ cofinite logarithmic conformal field theories.

\subsection{The modular differential equation} \label{ref:mdeq}

We will now show how the vacuum character determines the
spectrum of a conformal to a very high extent. In fact, a (sufficiently high)
finite order of the $q$-expansion of the vacuum character is enough, and can
often be computed by explicitly constructing a basis of states of the vacuum 
representation space up to this finite level. Beyond this, almost nothing else
needs to be known. The key is that in order to be a $C_2$ cofinite theory,
the vacuum character, as all other torus amplitudes,  must satisfy a modular 
invariant differential equation. The equation with the smallest possible degree
$n$ is uniquely determined by the vacuum character, as long as we know more
than the $n$ first terms of it. If we do not know $n$, i.e. if we do not know
the dimension of the space of the torus amplitudes (in case of a rational
conformal field theory, this is just the number of inequivalent highest
weight representations), we might be able to estimate it, if we can compute
the dimensions of the graded subspaces of the vacuum representation up to some
(possibly large) level.

We will make use of the prime example of logarithmic conformal field theory,
the $c=-2$ triplet model, in order to demonstrate this approach.

\subsubsection{The $c=-2$ triplet theory} \label{ref:triplet}

Let us briefly recall some of the properties of the triplet theory with
$c=c_{2,1}=-2$, which can be found e.g. in
\cite{Kau91,Flo96,Kau95,GK96b}. The chiral algebra for this
conformal field theory is generated by the Virasoro modes $L_n$, and
the modes of a triplet of weight 3 fields $W^a_n$. The commutation
relations are   
\begin{subequations}
\begin{align}
  {}[ L_m, L_n ] =& (m-n)L_{m+n} - \frac16 m(m^2-1) \delta_{m+n}, 
  \\
  {}[ L_m, W^a_n ] =& (2m-n) W^a_{m+n}, 
  \\
  {}[ W^a_m, W^b_n ] =& g^{ab} \biggl( 
  2(m-n) \Lambda_{m+n} 
  +\frac{1}{20} (m-n)(2m^2+2n^2-mn-8) L_{m+n} 
  \nonumber\\
  &\qquad
  -\frac{1}{120} m(m^2-1)(m^2-4)\delta_{m+n}
  \biggr) 
  \nonumber \\
  &
  + f^{ab}_c \left( \frac{5}{14} (2m^2+2n^2-3mn-4)W^c_{m+n} 
    + \frac{12}{5} V^c_{m+n} \right) 
  \end{align}
\end{subequations}
where $\Lambda = \mathopen:L^2\mathclose: - 3/10\, \partial^2L$ and 
$V^a = \mathopen:LW^a\mathclose: - 3/14\, \partial^2W^a$ are
quasiprimary normal ordered fields. $g^{ab}$ and $f^{ab}_c$ are the
metric and structure constants of $su(2)$. In an orthonormal basis we
have $g^{ab} = \delta^{ab}, f^{ab}_c = i\epsilon^{abc}$. 

The triplet algebra (at $c=-2$) is only associative, because certain
states in the vacuum representation (which would generically violate
associativity) are null. The relevant null vectors are
\begin{align} 
  N^a =& 
  \left(2 L_{-3} W^a_{-3} -\frac43 L_{-2} W^a_{-4} + W^a_{-6}\right)
  \Omega, 
\label{eq:nulllw}
\\
  N^{ab} =&
  W^a_{-3} W^b_{-3} \Omega - 
  g^{ab} \left( \frac89 L_{-2}^3 + \frac{19}{36} L_{-3}^2 +
  \frac{14}{9} L_{-4}L_{-2} - \frac{16}{9} L_{-6} \right)\Omega 
  \nonumber\\*&
  - f^{ab}_c\left( -2 L_{-2} W^c_{-4} + \frac54 W^c_{-6}
  \right) \Omega \,.
\label{eq:nullww}
\end{align}
We shall only be interested in representations which respect these
relations, and for which the spectrum of $L_0$ is bounded from below.
Evaluating the constraint coming from (\ref{eq:nullww}), we find (see
\cite{GK96b} for more details) 
\begin{equation}
  \left(W^a_0 W^b_0 - g^{ab} \frac19 L_0^2 (8L_0 + 1) - 
  f^{ab}_c \frac15 (6L_0-1) W^c_0 \right) \psi = 0 \,,
\label{eq:wwzero}
\end{equation}
where $\psi$ is any highest weight state, while the relation coming
from the zero mode of (\ref{eq:nulllw}) is satisfied identically. 
Furthermore, the constraint from $W^a_1 N^{bc}_{-1}$, together with
(\ref{eq:wwzero}) implies that $W^a_0 (8 L_0 - 3) (L_0 -1) \psi = 0$. 
Multiplying with $W_0^a$ and using (\ref{eq:wwzero}) again, this
implies that 
\begin{equation}
\label{eq:heigen}
0  = L_0^2 (8 L_0 + 1) (8 L_0 - 3) (L_0 - 1) \psi\,.
\end{equation}
For irreducible representations, $L_0$ has to take a fixed value $h$
on the highest weight states, and (\ref{eq:heigen}) then implies that
$h$ has to be either $h=0, -1/8, 3/8$ or $h=1$. However, it also
follows from (\ref{eq:heigen}) that a logarithmic highest weight 
representation is allowed since we only have to have that $L_0^2=0$
but not necessarily that $L_0=0$. Thus, in particular, a
two-dimensional space of highest weight states with relations
\be\label{3.16}
L_0\, \omega = \Omega \qquad \qquad L_0\, \Omega = 0 \,.
\ee
satisfies (\ref{eq:heigen}). This highest weight space gives rise to
the ``logarithmic'' (indecomposable) representation $\Rc_0$ (see
\cite{GK96b} for more details). Note that quite some effort is needed to
find the relation (\ref{eq:heigen}), as the ${\cal W}$-algebra null
vectors must be computed explicitly. As we will see shortly, we can derive
the same relation solely from the knowledge of the vacuum character, which can
be found without explicitly knowing the null vectors. In fact, as shown in
\cite{CF06}, the vacuum characters as well as $C_2$ cofiniteness of the whole
series of $c_{p,1}$ triplet models can be established without explicit 
knowledge of the triplet ${\cal W}$-algebra null vectors, which, for larger
$p$ would be impossible to get by.

It follows from the above analysis (and a similar analysis for the
$W^a$ modes; see for example \cite{GK96b}) that the $c=-2$ triplet theory
has only finitely many indecomposable highest weight
representations. This suggests that it satisfies the $C_2$ condition,
and this can be confirmed by a computer calculation (first done by Horst
Kausch, unpublished).
Indeed, the space $\Hi_0 / C_2(\Hi_0)$ has dimension
$11$, and it can be taken to be spanned by the vectors
\begin{align}
 L_{-2}^s\Omega\,,   & \qquad\hbox{where $s=0,1,2,3,4$} \nonumber \\
 L_{-2}^s W^a_{-3}\Omega \,, & \qquad\hbox{where $s=0,1$ and
$a\in$ adj(su$(2))$}\,.
\end{align}

\subsubsection{The modular differential equation} \label{ref:mdeq2}

The above calculation leading to (\ref{eq:heigen}) implies that in
Zhu's algebra we have the relation 
\be\label{triprel}
L_0^2 (8 L_0 + 1) (8 L_0 - 3) (L_0 - 1) = 0 \,,
\ee
where $L_0$ denotes the operator corresponding to the stress energy
tensor, and the product is to be understood as the product in Zhu's
algebra (see for example \cite{GG00} for an explanation of this
construction). Given (\ref{diff}) and (\ref{rel}) this therefore
suggests that there should be a fifth order differential equation that 
characterizes the vacuum torus amplitudes for the triplet theory, and
furthermore, that the leading order (\ref{rel}) should precisely
reduce to (\ref{triprel}).\footnote{Strictly speaking, the above
argument only implies that there should be a differential equation
characterizing the vacuum torus amplitudes whose order is at least
five. We shall assume in the following that the order is precisely five;
as we shall see later on this assumption leads to a consistent
description of the vacuum torus amplitudes.} Furthermore, since the
space of vacuum torus amplitudes is invariant under the action of the
modular group $\mathrm{PSL}(2,\Zop)$, the differential equation
must be modular invariant as well. The most general modular invariant
differential equation of degree five is 
\be\label{ansatz}
\left[ D^5 + \sum_{r=0}^{4} f_r(q)\, D^r \right] T(q) = 0 \,,
\ee
where each $f_r(q)$ is a polynomial in $E_4(q)$ and $E_6(q)$ of
modular weight $10-2r$, and 
\begin{equation}
D^i = {\it cod}_{(2i)}\cdots{\it cod}_{(2)} {\it cod}_{(0)}\,,
\end{equation}
with $cod_{s}$ being the modular covariant derivative on weight $s$ 
modular functions
\begin{equation}
{\it cod}_{(s)} = q {\partial \over \partial q}
		- \frac{1}{12} (s-2) E_2(q)\,,
\end{equation}
which increments the weight of a modular form by 2. Here $E_2$ is the
second Eisenstein series, and ${\it cod}_{(0)}f=f$. A differential
equation of this type can always be found if the space of vacuum torus
amplitudes forms a finite dimensional representation of the modular
group. For the case of rational conformal field theories, this
differential equation was first considered by \cite{MMS1,MMS2} (see also
\cite{Eho95,ES95} for further developments). It is often called the
{\it modular differential equation}.  

The first few of the $D^i$ read to first order in $q$, i.e. where 
$E_2(q)$  is only taken as $1-24q+{\cal O}(q^2)$, and with the
notation  $D_q=q\frac{\partial}{\partial q}$, simply
\begin{align}\nn
  D^0 =& 1\,,\\
\nn  D^1 =& D_q^{}\,,\\
\nn  D^2 =& D_q^2 - \frac16 D_q + q\,4D_q\,,\\
\nn  D^3 =& D_q^3 - \frac12 D_q^2 + \frac{1}{18}D_q
       + q\,\left(12D_q^2+\frac43 D_q\right)\,,\\
\nn  D^4 =& D_q^4 - D_q^3 + \frac{11}{36}D_q^2 - \frac{1}{36}D_q
       + q\,\left(24D_q^3 +\frac43D_q^2+\frac43D_q\right)\,,\\
  D^5 =& D_q^5 - \frac53 D_q^4 + \frac{35}{36}D_q^3 - \frac{25}{108}D_q^2 
                + \frac{1}{54}D_q
       + q\,\left(40D_q^4-\frac{20}{3}D_q^3+\frac{20}{3}D_q^2\right)\,,
\end{align}
where all expressions are up to ${\cal O}(q^2)$. Of course, $D^0$ and
$D^1$ are exact to all orders. 

The most general ansatz for the differential equation (\ref{ansatz})
is therefore 
\be\label{ansatz1}
\sum_{k=0}^5\sum_{{r,s\atop4r+6s=10-2k}} 
    a_{r,s}(E_4)^r(E_6)^s\left(
	\prod_{m=0}^{k}{\it cod}_{(2m)}\right) T(q)=0\,.
\ee
This differential equation must reduce to (\ref{triprel}) (in the
sense of (\ref{diff}) and (\ref{rel})) as $q\rightarrow 0$, and it
must furthermore be satisfied for the characters of the irreducible
highest weight representations of the triplet algebra. As we have
explained before, there are four irreducible highest weight
representations with conformal weights $h=0, -1/8, 3/8$ and $h=1$, and
their corresponding characters are known \cite{Flo96,Kau95,Flo97}. 
In terms of the functions of section \ref{ref:theta},
they are given as 
\begin{align}
\chi_{-{1\over 8}}(q) &= \theta_{0,2}(q)/\eta(q)\,,\label{car1}\\
\chi_0(q) &= (\theta_{1,2}(q)+(\partial\theta)_{1,2}(q))/\eta(q)\,,
\label{car2}\\
\chi_{{3\over 8}}(q) &= \theta_{2,2}(q)/\eta(q)\,,\label{car3}\\
\chi_1(q) &= (\theta_{1,2}(q)-(\partial\theta)_{1,2}(q))/\eta(q)\,.
\label{car4}
\end{align}
Putting these pieces of information together we find that (up to an
overall normalisation constant) (\ref{ansatz1}) is uniquely determined
to be the differential equation
\begin{align}
0 =  & \left[\frac{143}{995328}E_4(q) E_6(q) +
   \frac{121}{82944}(E_4(q) )^2{\it cod}_{(2)} +
    \frac{65}{2304}E_6(q) {\it cod}_{(4)}{\it cod}_{(2)}
    \right.\nonumber\\
&
\left.\mbox{}-\frac{163}{576}E_4(q) {\it cod}_{(6)}{\it cod}_{(4)}
			{\it cod}_{(2)} +
	{\it cod}_{(10)}{\it cod}_{(8)}{\it cod}_{(6)}{\it cod}_{(4)}
			{\it cod}_{(2)}\right]T(q)\,.
\end{align}
It is instructive to look at the leading order of the above equation.
If we expand the Eisenstein series $E_n = 1 + g_{n,1}q + {\cal O}(q^2)$ 
with $g_{n,1}$ given by $g_{2,1}=-24,g_{4,1}=240,g_{6,1}=-504$, we
obtain 
\begin{align}
\nn  0 =&
  \left(D_q^5 - \frac53 D_q^4 + \frac{397}{576} D_q^3 -
  \frac{427}{6912}D_q^2    -\frac{37}{82944}D_q +
  \frac{143}{995328}\right)T(q) \\ 
    &+ q\left(40 D_q^4 - \frac{895}{12}D_q^3  + \frac{2209}{96}D_q^2
    - \frac{209}{216}D_q - \frac{1573}{41472}\right)T(q)
    + {\cal O}(q^2)\,.
\end{align}
The zero-order term in $q$ can be factorized as
\be\label{lowest}
  \frac{1}{995328}(24D_q-11)(12D_q-13)(24D_q+1)(12D_q-1)^2\,.
\ee
Recalling that $D_q$ has to be replaced by 
$L_0-{c\over 24}=L_0+{1\over 12}$ in order to relate
(\ref{diff}) to (\ref{rel}), this therefore reduces, as required,
to (\ref{triprel}). If we make the ansatz
\be\label{torusans}
T(q) = q^{h+{1\over 12}} \left( 1 + c_1 q + c_2 q^2 + c_3 q^3 + 
{\cal O}(q^4) \right)\,,
\ee
the above differential equation becomes, up to third order,
\begin{align}
  0=&\frac{q^{h+1/12}}{64}\left[q^0\left(h^2(h-1)(8h+1)
                                (8h-3)\right)\right.\nn\\
   &+ q^1\left(c_1(h+1)^2h(8h+9)(8h+5) 
             + 2h(32h-45)(40h^2-5h-1)\right)\nn\\
   &+ q^2\left(c_2(h+2)^2(h+1)(8h+17)(8h+13) \right.\nn\\
             &\quad+ 2c_1(32h-13)(h+1)(40h^2+75h+34)\nn\\
   &  \left.\quad+ 2(3840h^4+2840h^3-17331h^2+706h-442)\right)\nn\\
   &+ q^3\left(c_3(h+3)^2(h+2)(8h+25)(8h+21) \right.\nn\\
&\quad             + 2c_2(h+2)(32h+19)(40h^2+155h+149)\nn\\
   &  \left.\quad+ 2c_1(3840q^4+18200q^3
                       +14229q^2-10076q-10387)\right.\nn\\
   &  \left.\left.\quad+ 4(2560h^4+28880h^3
                       -66574h^2-9772h-12281)\right)
       + {\cal O}(q^4)\right]\,.
\end{align}

\subsubsection{Solving the modular differential equation} \label{ref:solmdeq}

As we have seen, the modular differential equation is of fifth order
for the triplet theory, and the space of vacuum torus amplitudes is
therefore five-dimensional. On the other hand, we have only got four
irreducible representations that give rise, via their characters, to
four vacuum torus amplitudes (that solve the differential
equation). Let us now analyze how to obtain a fifth, linearly
independent, vacuum torus amplitude. First let us try to find a
solution of the form (\ref{torusans}). Because of the lowest order
equation (\ref{lowest}), this will only give rise to a solution
provided that $h=-{1\over 8}, {3\over 8}, 0$ or $h=1$. For each fixed
$h$, one then finds that there is only one such solution, which
therefore agrees with the corresponding character of the irreducible 
representation (i.e. with (\ref{car1}) -- (\ref{car4})). By the way,
this conclusion was not automatic {\it a priori}, since there exist
cases where the modular differential equation has two linearly
independent solutions with the same conformal weight, both of which
are of power series form. The simplest example is provided by the two
$h=0$ characters of the $c=1-24k$ series of rational CFTs, $k\in\Nop$,
with extended symmetry algebra ${\cal W}(2,8k)$. One of these
solutions belongs to the vacuum representation, the other to a second
$h=0$ representation which, however, has a non-vanishing 
eigenvalue $w$ of the $W_0$ zero mode \cite{Flo93}.   

The character of any highest weight representation always gives rise
to a torus amplitude as in (\ref{torusans}), and thus we have shown
that {\it the space of vacuum torus amplitudes for the triplet theory
is not spanned by the characters of the (irreducible) highest weight
representations}. In fact, we find that the missing, linearly
independent solution can be taken to be 
\begin{equation}
T_5(q) = \log(q)(\partial\theta)_{1,2}(q)/\eta(q)\,.
\end{equation}
It is tempting to associate this vacuum torus amplitude with the
logarithmic (indecomposable) highest weight representation $\Rc_0$ 
whose ground state conformal weight is $h=0$ \cite{Flo96}. However,
as we have just explained, this identification can only be
formal. Furthermore, $T_5(q)$ is not uniquely determined by the above
considerations, since we can (obviously) replace $T_5(q)$ by  
$T'_5(q) = T_5(q)+\alpha_0 \chi_0(q) + \alpha_1 \chi_1(q)$ for any (real) 
$\alpha_i$, $i=1,2$. Additionally, it is not clear how $T_5(q)$ can be connected
to the character of the logarithmic representation $\Rc_0$, 
as $T_5(q)$ is not even a $q$-series.

Moreover, it is not difficult to compute the full character of the
indecomposable representation with $h=0$, and it turns out that it
is linearly dependent to the solutions we already have found. In fact,
\begin{equation}
	\chi_{{\cal R}_0} = \chi_0(q)+\chi_1(q) = 2\theta_{1,2}(q)/\eta(q)\,.
\end{equation}
To summarize, the modular differential equation has provided us with
four linear independent solutions which can all be interpreted as 
characters of irreducible representations. The characters of the two
indecomposable representations ${\cal R}_0\cong{\cal R}_1$ turn out
to be linear combinations of these four solutions.

It therefore seems that, unlike the case of a rational conformal field
theory where a (canonical) basis for the space of vacuum torus
amplitudes is given in terms of the characters of the irreducible
representations, the space of vacuum torus amplitudes does not possess
a canonical basis in our case. In particular, Verlinde's formula
therefore cannot make sense since it presupposes such a canonical
basis. (The Verlinde formula involves the matrix elements of the
$S$-modular transformation; these matrix elements are only defined
once a basis for the space of torus amplitudes has been chosen.) This is
in nice agreement with the fact that the fusion rules of the triplet
theory cannot be diagonalized \cite{GK96b}, and therefore that they
cannot be described by a Verlinde formula. 

However, there exists a generalization of the Verlinde formula which works on
the space of all torus amplitudes and then projects out all contributions from
torus amplitudes which are not contained in the space of characters by a 
limit procedures {\slshape after \/} the fusion rules have been computed
\cite{Flo97}. The equivalence of the fusion rules computed via this 
generalized Verlinde formula with the correct fusion rules obtained by more 
direct methods (e.g. \cite{FFHST02}) is shown in \cite{FK07}.

\subsubsection{A more general analysis} \label{ref:genmdeq}

For any (logarithmic) conformal field theory which satisfies the $C_2$
condition, the mere existence of a finite order differential equation
allows us to derive some relations and bounds for the highest weights.
As argued in the previous sections for the special case of the $c=-2$ triplet
algebra, the torus amplitudes of such a theory have to satisfy a
$n$-th order holomorphic modular invariant differential equation of
the form (\ref{ansatz}), 
\be\label{ansatz2}
\left[D^n + \sum_{r=0}^{n-1}f_r(q)D^r\right]T(q) = 0\,,
\ee
where the $f_r(q)\in\Cop[E_4,E_6]$ are modular functions of weight
$2(n-r)$. These coefficient functions may be expressed in terms of a
set of $n$ linearly independent solutions $T_1(q),\ldots,T_n(q)$ of
the differential equation (\ref{ansatz2}). Note, however, that in
contrast to \cite{MMS1,MMS2}, these solutions cannot in general be
identified with the characters of representations. In particular, we
cannot therefore assume that the $T_i(q)$ have a good power series
expansion in $q$ up to a common fractional power 
$h_i - c/24\ {\rm mod}\ 1$.\footnote{We will in the following always
speak of power series expansions in $q$ with the silent understanding
that a common fractional power is allowed, i.e. that 
the functions can be expanded as 
$T(q)=q^\alpha\sum_{k=0}^{\infty} a_kq^k$, $\alpha\in\mathbb{Q}$.}
As we have seen, in logarithmic conformal field theories torus
amplitudes are not all elements in $\Cop(\!(q)\!)$, but may be in
$\Cop(\!(q)\!)[\tau]$, {\em i.e.} they are power series in $q$ times a
polynomial in $\tau\equiv{1\over 2\pi{\rm i}}\log(q)$. 

The following analysis along the lines set out in \cite{MMS1,MMS2} has to take
into account this fact. Therefore, we will {\em not\/} assume in the
following that the highest weights are all different, $h_i\neq h_j$ for
$i\neq j$, but only that $T_i(q)\neq T_j(q)$ for $i\neq j$. Note that
the asymptotic behavior of two functions $T_i(q)$ and $T_j(q)$ in the
limit $q\rightarrow 0$ (or $\tau\rightarrow+{\rm i}\infty$) is the same
whenever $T_j(q)=p(\tau)T_i(q)$ for a polynomial $p$, provided 
$T_i(q)\sim q^\alpha$ with $\alpha\neq 0$. The case $\alpha=0$ occurs
precisely when $h_i-c/24=0$. It is interesting to note that all known
logarithmic conformal field theories, except for $c=0$, share the fact that 
there is at most
one irreducible representation with $h=c/24$. 
Our analysis suggests that this should be generally true 
such that we only need one function $T(q)$ with asymptotic behavior 
$T(q)\sim 1+{\cal O}(q)$ for $q\rightarrow 0$. We will treat the case
$c=0$ separately below. 

With this in mind, we can express the coefficients of the modular
differential equation in terms of the Wronskian of a set of $n$
linearly independent 
solutions as
\begin{align}
  f_r(q) &= (-1)^{n-r}W_r(q)/W_n(q)\,,\nonumber\\
  W_r(q) &= {\rm det}\left(\begin{array}{rcr}
         T_1(q) & \phantom{m}\ldots\phantom{m} &        T_n(q) \\
      D^1T_1(q) &            \ldots            &     D^1T_n(q) \\
    \vdots\phantom{m}  &                   & \vdots\phantom{m} \\
  D^{r-1}T_1(q) &            \ldots            & D^{r-1}T_n(q) \\
  D^{r+1}T_1(q) &            \ldots            & D^{r+1}T_n(q) \\
    \vdots\phantom{m}  &                   & \vdots\phantom{m} \\
      D^nT_1(q) &            \ldots            &     D^nT_n(q) 
  \end{array}\right)\,.
\end{align}

The fact that $f_r(q)\in\Cop[E_4,E_6]$ puts severe constraints on the
possible polynomials $p_i(\tau)$ occurring in
$T_i(q)=p_i(\tau)q^{h_i-{c\over 24}}(1 +a_{i,1}q^1+a_{i,2}q^2 +\ldots)$.  
However, the explicit investigation of these constraints is 
beyond the scope of the present paper. For us, it is enough to note
that known examples such as the torus amplitudes of the $c_{p,1}$
logarithmic conformal field theories (with $c_{2,1}=-2$ being the
triplet model considered so far) do yield the correct coefficient
functions in this way. The $c_{p,1}$ models, as all logarithmic
conformal field theories with indecomposable representations spanning
Jordan blocks of rank two, involve only torus amplitudes $T(q)$ whose
polynomial parts are either constant or at most of degree one in
$\tau$. In general, indecomposable representations with Jordan cells
of rank $R$ will involve torus amplitudes with polynomials in $\tau$
of degree $d<R$.

The torus amplitudes, considered as functions in $\tau$ are
non-singular in $\mathbb{H}$. As a consequence, the same applies for
the $W_r$. Therefore, the coefficients $f_r$ can have singularities
only at the zeroes of $W_n$.  We will see that the total number of
zeroes of $W_n$ can be expressed in terms of the number $n$ of
linearly independent torus amplitudes, the central charge $c$ and the
conformal weights $h_i$ associated to the torus amplitudes
$T_i(q)$. In order to do so, we note that in the 
$\tau\rightarrow+{\rm i} \infty$ limit, the torus amplitudes behave as
$\exp(2\pi{\rm i}(h_i - {c\over 24})\tau)$. With the above caveat 
concerning the case $h=c/24$, this applies to all 
torus amplitudes independently of whether they are pure power series
in $q$, or whether they have a $\tau$-polynomial as additional
factor. This implies that 
$W_n\sim\exp(2\pi{\rm i}(\sum_ih_i - n{c\over 24})\tau)$, which says
that $W_n$ has a pole of order $n{c\over 24}-\sum_ih_i$ at
$\tau={\rm i}\infty$.  Now, $W_n$ involves precisely ${1\over 2}n(n-1)$
derivatives meaning that it transforms as a modular form of weight
$n(n-1)$. Both facts together allow us to compute the total number of
zeroes of $W_n$, which is 
\be\label{zeroes}
  {1\over 6}\ell\equiv 
  -\sum_{i=1}^n h_i + {1\over 24}nc + {1\over 12}n(n-1)\geq 0\,.\qquad
  \ell\in\mathbb{Z}_+-\{1\}\,.
\ee
This number cannot be negative since $W_n$ must not have a pole in the
interior of moduli space. We note that (\ref{zeroes}) is always a
multiple of ${1\over 6}$ since $W_n$, as a single valued function in
Teichm\"uller space, may have zeroes at the ramification points
$\exp({1\over 3}\pi{\rm i})$ and $\exp({1\over 2}\pi{\rm i})$ of order
${1\over 3}$ and ${1\over 2}$, respectively. Equation (\ref{zeroes})
provides a simple bound on the sum of the conformal weights. 

For example, for the case of the $c=-2$ triplet theory, we have 
\be
  -\left[(-{1\over 8})+(0)+(0)+({3\over 8})+(1)\right] 
      + {1\over24}(5)(-2) + {1\over 12}(5)(4) = 0\,.
\ee
This is now sufficient to exclude that the $c=-2$ triplet theory 
is governed by a modular differential equation of order six. For if it
was of order six, the additional sixth conformal weight would have to
be of the form $h={3\over 4} - {1\over 6}\ell$. This is obviously
always a rational number with denominator either $4$ or $12$. However,
this weight does not belong to the list of admissible weights of the
$c=-2$ triplet theory, even not modulo one, and thus we have a
contradiction. One can also check this directly: assuming that we
have a modular differential equation of order six and starting with
the four irreducible characters (\ref{car1}) -- (\ref{car4}) one
obtains an ansatz for the modular differential equation depending on
the choice of the  conformal weights of the two remaining
representations. One finds that no consistent choice is
possible, and thus no six-dimensional representation of the modular
group containing the characters (\ref{car1}) -- (\ref{car4}) can be
found.  

\subsubsection{The other triplet theories} 

The analysis presented so
far  can in principle be generalized to all members of the $c_{p,1}$
series of triplet models. In practice, however, we have not found it
possible to give uniform explicit expressions.
The pattern which emerges in the treatment of the $c=-2$ case, i.e.
the case $p=2$, however, seems to be of a generic nature. Indeed, all
the $c_{p,1}$ models are $C_2$ cofinite \cite{CF06} and the characters
of their irreducible representations are all known. They close under
modular transformations provided that a certain number of ``logarithmic
vacuum torus amplitudes'' (the analogues of $T_5(q)$) are added to the
set. In fact, the characters of the irreducible representation,
together with  additional torus amplitudes which we may again
associate to the indecomposable representations, read \cite{Flo96}  
\begin{align}
  \chi^{}_{0,p}(q)      & =  \frac{1}{\eta(q)}\Theta_{0,p}(q)\,,\\
  \chi^{}_{p,p}(q)      & =  \frac{1}{\eta(q)}\Theta_{p,p}(q)\,,\\
  \chi^+_{\lambda,p}(q) & =  \frac{1}{p\eta(q)}\left[
                            (p-\lambda)\Theta_{\lambda,p}(q)
                          + (\partial\Theta)_{\lambda,p}(q)\right]\,,\\
  \chi^-_{\lambda,p}(q) & =  \frac{1}{p\eta(q)}\left[
                            \lambda\Theta_{\lambda,p}(q)
                          - (\partial\Theta)_{\lambda,p}(q)\right]\,,\\
  \tilde{\chi}^{}_{\lambda,p}(q) & =  \frac{1}{\eta(q)}\left[
                            2\Theta_{\lambda,p}(q)
                            - \mathrm{i}\alpha\log(q)
                            (\partial\Theta)_{\lambda,p}(q)\right]
  \,,
\end{align}
where $0<\lambda<p$ and 
where we made use of the definitions (\ref{etadef}) to
(\ref{deltheta}). As before, the ``logarithmic'' torus amplitudes
$\tilde{\chi}_{\lambda,p}$ are not uniquely determined by these
considerations since $\alpha$ is a free constant; the form given above
is convenient for constructing modular invariant partition
functions. One should note, however, that for 
logarithmic conformal field theories the complete space of states of
the full non-chiral theory is not simply the direct sum of   
tensor products of chiral representations (see for example
\cite{GK99}). 
It is therefore not clear how the full
torus amplitude has to be constructed out of these generalized
characters.  

The congruence subgroup for the $c_{p,1}$ model is $\Gamma(2p)$. There
are $2p$ characters corresponding to irreducible representations, and
$(p-1)$  ``logarithmic'' torus amplitudes, giving rise to a $(3p-1)$
dimensional representation of the modular group. In particular, we
therefore expect that the order of the modular differential equation
is $(3p-1)$. Furthermore, we expect that the dimension of Zhu's
algebra is $6p-1$: it follows from the 
structure of the above vacuum torus amplitudes that $p$ of the
irreducible representations have a one-dimensional ground state space,
while the other $p$ irreducible representations have ground state
multiplicity two; as above one may furthermore expect that each of the
$(p-1)$ logarithmic representations probably leads to one additional
state, thus giving altogether the dimension $p + 4p + (p -1) = 6p-1$. 
In deed, this was proved in \cite{Adamovic:2010zk}. In addition, the same
authors obtained a modular differential equation of order $3p-1$ for all $p$, 
satisfied by all the torus amplitudes \cite{AdamovicMilas}. 

While it is not possible to write down a general expression for 
the modular differential equation for all $p$, we can give support for
these conjectures by analyzing the $p=3$ triplet model with
$c=-7$. The vacuum character of this theory is $\chi^+_{2,3}(q)$. 
Under the assumption that the modular differential equation
is in fact of order $3p-1=8$, we can determine it uniquely by
requiring it to be solved by this vacuum character. Explicitly, we find
\begin{align}
0 =& \left[
  \Big(\frac{833}{53747712}E_4(q)(E_6(q))^2
      -\frac{990437}{36691771392}(E_4(q))^4\Big)
  \right.\nonumber\\  & \left.\mbox{}
  -\frac{40091}{143327232}(E_4(q))^2E_6(q){\it cod}_{(2)}
  \right.\nonumber\\  & \left.\mbox{}
  +\Big(\frac{115}{746496}(E_6(q))^2
       +\frac{53467}{47775744}(E_4(q))^3\Big){\it cod}_{(4)}{\it cod}_{(2)}
  \right.\nonumber\\  & \left.\mbox{}
  -\frac{5897}{124416}E_4(q)E_6(q){\it cod}_{(6)}{\it cod}_{(4)}
                                  {\it cod}_{(2)}
  \right.\nonumber\\  & \left.\mbox{}
  +\frac{10889}{55296}(E_4(q))^2{\it cod}_{(8)}{\it cod}_{(6)}
                                {\it cod}_{(4)}{\it cod}_{(2)}
  \right.\nonumber\\  & \left.\mbox{}
  +\frac{157}{432}E_6(q){\it cod}_{(10)}{\it cod}_{(8)}
                        {\it cod}_{(6)}{\it cod}_{(4)}
                        {\it cod}_{(2)}
  \right.\nonumber\\  & \left.\mbox{}
  -\frac{21}{16}E_4(q){\it cod}_{(12)}{\it cod}_{(10)}
                      {\it cod}_{(8)}{\it cod}_{(6)}
                      {\it cod}_{(4)}{\it cod}_{(2)}
  \right.\nonumber\\  & \left.\mbox{}
  +{\it cod}_{(16)}{\it cod}_{(14)}{\it cod}_{(12)}{\it cod}_{(10)}
   {\it cod}_{(8)}{\it cod}_{(6)}{\it cod}_{(4)}{\it cod}_{(2)}
   \vphantom{\frac{1}{2}}\right]T(q)\,.
\end{align}
If we make the ansatz that $T(q)$ is of the form 
\be
T(q) = q^{h-\frac{c}{24}}\, \sum_{n=0}^{\infty} \, 
\sum_{k=0}^{1} c_{k,n} \tau^k \, q^m \,, 
\ee
we obtain, to lowest order the polynomial condition
\begin{align}
  0=& \frac{1}{2304}(1+4h)^2h^2(h-1)(12h-5)(3h+1)(4h-7)(c_{1,0}\tau
  +c_{0,0})
  \\
  +&\frac{1}{1152}(1+4h)h(2304h^5-5280h^4+2160h^3+870h^2-229h-35)c_{1,0}
  \ +\ {\cal O}(q) \,.\phantom{mmm}\nn
\end{align}
As expected, we can read off from this expression the 
allowed conformal weights: if the character does not involve any
powers of $\tau$ ($c_{1,0}=0$), then $h$ needs to be from the set 
$h\in\{0,-1/4,1,5/12,-1/3,7/4\}$. Furthermore, we have two
``logarithmic'' torus amplitudes with $h=0$ and $h=-1/4$. This then fits
nicely together with the fact that there are in fact two 
indecomposable highest weight representations with these conformal
weights \cite{GK96a}.

We see that the modular differential equation is a powerful tool to investigate
a specifically given conformal field theory where it may greatly help to
understand its representation theory from a comparable small amount of
pre-knowledge. However, it is less suitable to derive general statements 
about (certain series of) rational conformal field theories. But as we will
see in the next section, we can draw some general conclusions from the
rather restrictive conditions, the spectrum of a rational conformal field 
theory must satisfy, such that a modular differential equation can exist (which
it must, if the theory is indeed rational).

\subsubsection{Augmented minimal models} \label{ref:augminmod}

The diophantine equation (\ref{zeroes}) is very powerful. It allows
us to estimate whether a conformal field theory with given central
charge $c$ and a certain set $\{h_i:i\in I\}$ of known weights of
representations has a chance to be rational in the sense that the
associated torus amplitudes $T_i(q)$ can from a finite-dimensional
representation of the modular group. To illustrate this, let us
consider a conformal field theory with degenerate representations in
the sense of BPZ \cite{BPZ84}. Let us further, for the sake of
simplicity, restrict ourselves to the theories in the Kac table
\be
  c=c_{p,q}=1-6\frac{(p-q)^2}{pq}\,,\ \ \ \
  h_{r,s}(c_{p,q}) = \frac{(pr-qs)^2-(p-q)^2}{4pq}\,.
\ee
However, we do not at this stage restrict ourselves to minimal models,
so we do not  prescribe finite ranges for $r,s$. It is clear that
the standard (not truncated) fusion rules 
\be
  [r,s]*[r',s']=\sum_{{\rho=|r-r'|+1\atop r+r'+\rho\equiv 1\,\mathrm{mod}\,2}}
                    ^{r+r'-1}
               =\sum_{{\sigma=|s-s'|+1\atop s+s'+\sigma\equiv 1\,\mathrm{mod}\,
	            2}}^{s+s'-1}[\rho,\sigma]
\ee
imply that any possible finite subset of the infinite Kac table must be
a rectangle starting at $[1,1]$ and ending at $[b,a]$ such that
$1\leq r\leq b$, $1\leq s\leq a$. 
In the minimal models we have $a=p-1$ and $b=q-1$. Summing up all
these $n=a\cdot b$ weights in the formula (\ref{zeroes}) yields
\be
  -\frac{ab}{24pq}( (2b+1)p-(a+1)q )( (b+1)p - (2a+1)q )\,.
\ee
This generically can only be a solution of the form $\ell/6$, $\ell\in
\mathbb{Z}_+$, if $a=\alpha p-1$, $b=\beta q-1$ with
$\alpha,\beta\in\mathbb{N}$ or if $a=\alpha p$, $b=\beta q$. We discard the
second type of solution, since it generically leads to $\ell <0$.
Thus, the only possible finite subsets of the
Kac table are formally obtained as the conformal grids of $c_{\alpha p,\beta q}$
instead of $c_{p,q}$. Assuming this, yields
\be
  \frac{(\alpha p-1)(\beta q-1)}{24}(p(2\alpha-\beta)-1)(q(2\beta-\alpha)-1)
  \,.
\ee
A further constraint comes from the fact that we have naively counted all
weights in a sub-rectangle of the Kac table. This yields too many
representations, since $h_{r,s}(c_{p,q}) = h_{\delta q-r,\delta p-s}(c_{p,q})$
for all $\delta\in\mathbb{N}$.\footnote{
If we do not assume the representations to be irreps, we might be tempted to
relax from identifying the representations according to the symmetry of the
weights. For instance, this could be the case, if we do not presuppose that all
representations possess more than one null vector (as in the minimal models). 
Concretely, it is common in the literature to associate all the various
representations of the $(1,p)$ triplet models or of logarithmically extended
minimal models, irreps as well as indecomposable representations, with
entries from extended Kac-tables. see e.g.\ \cite{Flo03,EF06b}.
However, we still cannot neglect the symmetry of the weights, as it reflects, 
for example, in the fusion rules.}
This leaves us with the only
sensible case $\alpha=\beta$, which means that only multiples of the
standard conformal grid are possible. Thus, instead of the conformal grid
for $c_{p,q}$ there might exist a finite-dimensional representation of
the modular group with functions $T_i(q)$, whose asymptotics is determined
by an enlarged conformal grid, formally obtained by considering
$c_{\alpha p,\alpha q}$ instead of $c_{p,q}$, thus allowing for a grid
size determined by non-coprime numbers $p'=\alpha p$, $q'=\alpha q$.
We hence find the diophantine condition
\be
  \frac{\ell}{6} = \frac{1}{24}(\alpha p-1)^2(\alpha q-1)^2\,.
\ee
Therefore, either $\alpha p$ or $\alpha q$ must be odd in order to kill
a factor of at least four from the denominator. Since $p$ and $q$ are
assumed to be coprime, one of them bust be odd. Thus, the diophantine
condition can indeed be satisfied with an $\ell\in\mathbb{Z}_+-\{1\}$ for
all $\alpha$ odd.

There is still one subtlety. When we consider conformal grids of this form,
all weights within this grid appear at least twice. Thus we should attempt
to only take half of the weights. Assuming now from the beginning a 
conformal grid of size $(\alpha p-1)(\alpha q-1)$, $\alpha$ odd, 
but only taking half of the weights, we find that the condition (\ref{zeroes})
is satisfied with $\ell=0$. This is a particularly nice result, since it means
that the determinant $W_n$ has no zeroes in the interior of moduli
space. Thus there are no values $\tau$ or $q$, respectively, 
for which the torus amplitudes become linearly dependent. Taking this as
a guide line, one can do the analysis a bit more refined. 

In order to achieve $\ell=0$, we should take into account the symmetry 
$h_{r,s}=h_{\alpha q-r,\alpha p-s}$ right from the beginning. 
The diophantine condition then reduces to
\be
  {1\over 6}\ell\equiv
  -\frac{1}{48}\frac{ab[(a+1)q-(b+1)p][(2a+1)q-(2b+1)p]}{pq}
   \,.\qquad \ell\in\mathbb{Z}_+-\{1\}\,.
\ee
Obviously, the numerator has two series of non-trivial but generally valid 
solutions to make it vanish, namely $(a,b)=(\alpha p-1,\alpha q-1)$ and
$(a,b)=((\alpha p-1)/2,(\alpha q-1)/2)$. 
The second series is only valid when both,
$\alpha q-1$ and $\alpha p-1$ are even integers implying that $p$ and $q$ 
must both be
odd then. Here, $\alpha\in\mathbb{N}$, of course. This result suggests that
any extension of the Kac table by a common integer factor $\alpha$ 
could yield a 
set of conformal weights such that the diophantine condition is satisfied
and therefore a modular differential equation may exist. However, we have to be
a bit more careful here. When counting the number of representations 
given by the Kac table of size $a\cdot b$, one typically has to take only
one half of them, since the other half is naturally identified via the
Virasoro module embedding structure. If $a\cdot b$ is odd, then this cannot
be done without one remaining field which appears only once in the Kac table.
Thus, there is another condition such that our generic solutions might
work, namely
\be
  (\alpha p-1)(\alpha q-1)\in2\mathbb{N}\quad{\rm or}\quad 
  (\alpha p-1)(\alpha q-1)/4\in2\mathbb{N}\,.
\ee
The second solution is thus more restrictive then the first, so we can 
concentrate on the easier case. We see immediately that the case
$\alpha=2$ is ruled out as a possible solution since this automatically
produces an odd number. Therefore, the next possible case after the
generic solution $\alpha=1$ is $\alpha=3$. Note that $\alpha=1$ always works 
since $p,q$ are coprime by assumption and thus cannot be both even. In fact,
the case $\alpha=2\alpha'$ even does not solve the diophantine equation at 
all, if the number of representations is counted correctly. Namely, the
entry $h_{\alpha'q,\alpha'p}(c_{p,q})$ 
exactly in the center of the conformal grid appears only once.
Taking this into account properly leads to a different result. Instead
of $\ell=0$, we now find that
\be
  \ell=pq(p+q){\alpha'}^3-(pq)^2{\alpha'}^4+\frac{1}{4}\Big(1 - \alpha'\Big[
       p(\alpha'p+1)+q(\alpha'q+1)\Big]\Big)\,.
\ee
Since $\ell$ must be integer, we immediately see that $\alpha'$ must be odd.
But even then can we not solve this condition, since then
$n(\alpha'n+1)\equiv n(n+1) \equiv 0$ modulo $2$, such that there is no
chance to cancel the remaining $1/4$.

Summarizing, the diophantine equation implies that augmented minimal models
might exist, if the weights are taken from the enlarged Kac table of size 
$(\alpha p-1,\alpha q-1)$ for any odd $\alpha$. The logarithmic triplet
theories are an example, since the admissible representations are taken from
the conformal grid of $c=c_{3p,3}$, i.e. from the enlarged Kac table with
$\alpha=3$. Similarly, as shown in \cite{EF06a,EF06b} for the explicit cases
$c=c_{9,6}$ and $c_{15,6}$, logarithmic extensions of minimal models do
exist for $\alpha=3$. 

One should note that we can only manage $\ell=0$, since the result for
the appropriate sum of conformal weights from the Kac table typically
turns out to be a negative rational number. Thus solutions with $\ell>1$
are presumably not to be found within Virasoro minimal models.

Similar considerations could be conducted for generic, rational central
charges $c\neq c_{p,q}$ such that the entries from the Kac table are all 
rational. 

On the other hand, if for a given central charge we can compute the vacuum 
character of the chiral symmetry algebra, or at least a sufficiently high order 
of it, then we only need an estimate of the expected number of admissible
representations to check whether this is compatible with the given 
vacuum character and to identify the possible conformal weights. In principle,
one could attempt to set up modular differential equations for the case
$\ell=0$ and of degree $n$, $n=1,2,3,\ldots$, increasing $n$ until the vacuum 
character turns out to be a solution without contradictions. However, although
the choice $\ell=0$ is plausible, determining the degree of the modular 
differential equation without any further knowledge proves difficult even for
today's computers as soon as $n$ gets large.

Finally, for augmented minimal models with $c=c_{3p,3q}$, we do have some
knowledge about the vacuum character and a good guess about the degree of
the modular differential equation is satisfies, as well as of the asymptotics
of the other torus amplitudes (i.e. the conformal weights from the augmented
Kac table). From this we at least get a set of modular functions which span the
space of torus amplitudes. One may than exploit the strategy of \cite{Flo97} to 
find the correct linear combinations of these functions which form the actual
characters of the augmented minimal model. This works by seeking such
linear combinations of the basis of torus amplitude which bring the $S$-matrix
into a form suitable for the computation of fusion rules via the limit 
procedure of the generalized Verlinde formula proposed in that paper.

\subsubsection{The case $c=0$} \label{ref:c=0}

A particularly interesting case is a conformal field theory of vanishing
central charge. There are some highly interesting problems in 
two dimensions, especially percolation, where conformal field theory with
central charge $c=0$ seems to play an important role. 
In this case, the critical value for a conformal weight,
such that the asymptotic behavior of amplitudes is not independent of
a polynomial factor $p(\tau)$, is just $h=0$. Standard lore in logarithmic
conformal field theory, however, holds that the vacuum representation is
always part of an indecomposable structure. Moreover, the extended Kac table
for $c_{2,3}=0$ contains several entries with vanishing conformal weight.

It is therefore tempting to apply the technique of the modular differential
equation to this case and see whether we can relax the condition that
a torus amplitude $T(q)=p(\tau)q^{h-c/24}\sum_{k=0}^\infty(a_kq^k)$, where
$a_0\neq 0$, must have $p(\tau)\equiv 1$ for $h=c/24$. A conformal
field theory with vanishing central charge is a difficult object.
In case of unitarity, a vanishing central charge implies that the theory
is trivial, {\em i.e.}\ its field content is given by the identity field 
alone. The second possibility is, as known from string theory, that a unitary
theory with $c>0$ is tensored with a ghost theory with central charge $-c$,
such that the total central charge vanishes. However, we might be interested
in non-trivial (and therefore non-unitary) theories with vanishing central
charge, which cannot be decomposed into factors with non-zero central 
charges.

We claim that the Virasoro minimal model with
central charge $c=0$, which is trivial, can be extended to a non-unitary
logarithmic model with finitely many representations. According to the
general reasoning of the previous section (see also \cite{Flo96}), 
we expect that these representations are labelled by
the formal conformal grid one obtains by considering $c_{3p,3q}$ instead
of $c_{p,q}$. We know that the conformal weights from this enlarged 
Kac table do satisfy all diophantine conditions necessary to set up a
modular differential equation. Let us therefore assume that
there exists a rational, possibly logarithmic, conformal
field theory such that its characters must satisfy a modular differential 
equation, generalized to the case of logarithmic conformal field theories 
with finitely many representations. We may then expect that this theory 
also fulfills Zhu's $C_2$ condition. Again, 
not all solutions can be interpreted as characters
i.e. traces over irreducible modules, but they should be understood
as vacuum amplitudes on the torus. 

Thus, in the case $c_{2,3}=c_{6,9}$ at hand, the extended conformal grid
has size $5\cdot 8=40$ and thus the modular differential equation is of
order $40/2=20$, its solutions  
span the space of potential vacuum amplitudes of this theory.
Whether this 20-dimensional space and the related conformal field theory
have something to do with the problem of percolation will be investigated
in future work. The equation itself is extremely lengthy, so we give it
here by referring to its general form (\ref{ansatz1}) and merely listing the
coefficients:
\begin{align}
	a_{4,1}  =& -3857/144\,,\nn\\
\nn	a_{6,0}  =& 27455/1728\,,\\
\nn	a_{8,2}  =& 11448089/55296\,,\\
\nn	a_{10,1} =& -81132109/331776\,,\\
\nn	a_{12,0} =& 287917355/23887872\,,\\
\nn	a_{12,3} =& -24118392235/47775744\,,\\
\nn	a_{14,2} =& 2145063995/2359296\,,\\
\nn	a_{16,1} =& -1007439963335/6879707136\,,\\
\nn	a_{16,4} =& 16235170093025/110075314176\,,\\
\nn	a_{18,0} =& -7342258067105/82556485632\,,\\
\nn	a_{18,3} =& -242834836836605/330225942528\,,\\
\nn	a_{20,2} =& 422788707478865/2641807540224\,,\\
\nn	a_{20,5} =& 124711173611453/293534171136\,,\\
\nn	a_{22,1} =& 7535975797080575/47552535724032\,,\\
\nn	a_{22,4} =& -1070013044796865/5283615080448\,,\\
\nn	a_{24,0} =& -59176258416552175/2282521714753536\,,\\
\nn	a_{24,3} =& 93314657139863735/1141260857376768\,,\\
\nn	a_{24,6} =& -429817670835475919/2282521714753536\,,\\
\nn	a_{26,2} =& -4260482854319725/71328803586048\,,\\
\nn	a_{26,5} =& 2786843679881165/23776267862016\,,\\
\nn	a_{28,1} =& 3016483131491075/41085390865563648\,,\\
\nn	a_{28,4} =& -266846149383639005/20542695432781824\,,\\
\nn	a_{28,7} =& 97023065226363863/41085390865563648\,,\\
\nn	a_{30,0} =& 291326860458440185/246512345193381888\,,\\
\nn	a_{30,3} =& 458410214497514285/123256172596690944\,,\\
\nn	a_{30,6} =& -657895207190269363/246512345193381888\,,\\
\nn	a_{32,2} =& 1039516789897710575/986049380773527552\,,\\
\nn	a_{32,5} =& -56214246975189865/18260173718028288\,,\\
\nn	a_{32,8} =& 1735787603351876711/986049380773527552\,,\\
\nn	a_{34,1} =& -310946513309707165/2958148142320582656\,,\\
\nn	a_{34,4} =& -941732882831629265/4437222213480873984\,,\\
\nn	a_{34,7} =& 1099798540316599171/2958148142320582656\,,\\
\nn	a_{36,0} =& 17980722135844325/35497777707846991872\,,\\
\nn	a_{36,3} =& 437118904830824975/53246666561770487808\,,\\
\nn	a_{36,6} =& 589451531309053765/35497777707846991872\,,\\
\nn	a_{36,9} =& -26032177604788465/986049380773527552\,,\\
\nn	a_{38,2} =& -138927688403253475/638959998741245853696\,,\\ 
\nn	a_{38,5} =& 99983800440166775/35497777707846991872\,,\\
\nn	a_{38,8} =& -57253337340701425/23665185138564661248\,,\\
	a_{40,1+3k} =& 0\,,\ \ k=0,1,2,3\,.
\end{align}
Many of its solutions will have
the same asymptotic behavior modulo one, i.e.
$T_i(q) \sim T_j(q)$ mod $1$ for 
$T_i(q) = p_i(\tau)q^{h_i-c/24}(1+{\cal O}(q))$ and
$T_j(q) = p_j(\tau)q^{h_j-c/24}(1+{\cal O}(q))$, if $h_i-h_j\in\mathbb{Z}$.
Thus, the modular differential equation only gives us a basis in the
space of vacuum amplitudes, but cannot tell us what the physically
correct linear combinations are. What we get is therefore only a
20-dimensional set of modular forms out of which the characters of the
irreducible and indecomposable representations as well as some further 
functions, which can only be interpreted as vacuum amplitudes, should be
constructed.

Our modular forms are all constructed from 
the building blocks $ \eta(q)$, $\Theta_{\lambda,k}(q)$, $(\partial\Theta)_{\lambda,k}(q)$ defined in section \ref{ref:theta}, plus
\begin{align}
  (\nabla\Theta)_{\lambda,k}(q) =& \log(q)
    \sum_{n\in\mathbb{Z}}(2kn+\lambda)q^{\frac{1}{4k}(2kn+\lambda)^2}\,,\\
  E_r(q)                          =&
    1 - \frac{2r}{B_r}\sum_{n=0}^{\infty}\sigma_{r-1}(n)q^n\,,
\end{align}
with $\lambda,k\in\mathbb{Z}_+$ and $r>0$ even, and where $\sigma_p(n)$ is the sum of the $p$-th powers of all divisors of $n$ and
$B_r$ is the $r$-th Bernoulli number. Of course, the $E_r$ are the 
Eisenstein series. Note, however, that $E_2$ is actually not a good
modular form, because it has a cusp. However,
$E_2$ plays an important role as it appears, for example, in the modular
covariant derivative.

Let us now study the special case of $c=c_{2,3}=0$. Besides the trivial 
solution,
where the spectrum contains only the identity operator, the next smallest
modular differential equation one can write down for this central charge
and conformal weights from the Kac table has order 20. There is no equation
of smaller order where the corresponding diophantine conditions on the
central charge and the conformal weights can be satisfied with weights taken
from the Kac table. The solutions of this equation factorize into three
sets which are separately closed under modular transformations. However,
the character of the irreducible vacuum representation (which is part of
a larger indecomposable representation), is presumably a linear combination 
of some of the 20 solutions in such a way that these three independent sets 
are intermixed. It is not yet known what the precise
vacuum character is, but in a work on alternating sign matrices by
P.~Pearce, {\em el al.}, two different characters for
two inequivalent $h=0$ representations of a $c=0$ Virasoro model are 
computed up to level five \cite{Pearce01}. Taking these as ansatz supports our
conjectures that firstly, there is more than one $h=0$ representation, and
that secondly, the characters of the $h=0$ representations are non-trivial
linear combinations of our solution set intermixing the three factors.
The three factors of the solution are of dimension six, ten and four,
respectively. They read:
\begin{align}
\nn  {\cal M}_1 =& {\rm span}\left\langle
                 \frac{1}{\eta(q)}(\Theta_{1,6}(q)+\Theta_{5,6}(q))\,,\ \
                 \frac{1}{\eta(q)}\Theta_{\lambda,6}(q) : 
		 \lambda\in\{0,2,3,4,6\}
                 \right\rangle\,,\\
\nn  {\cal M}_2 =& {\rm span}\left\langle
                 \frac{1}{\eta(q)}(\partial\Theta)_{\lambda,6}(q) :
		 \lambda\in\{1,2,3,4,5\}\,,\ \
                 \frac{1}{\eta(q)}(\nabla\Theta)_{\lambda,6}(q) :
		 \lambda\in\{1,2,3,4,5\}
                 \right\rangle\,,\\
  {\cal M}_3 =& {\rm span}\left\langle
                 \frac{1}{\eta(q)}(\Theta_{1,6}(q)-\Theta_{5,6}(q))\,,\ \ 
\nn		 E_2(q)\,,\ \ \log(q)E_2(q)\,,
		 \right.\\
	      & \phantom{{\rm span}}\,\ \ \left.
		 \log^2(q)E_2(q) + 12\log(q)
                 \frac{1}{\eta(q)}(\Theta_{1,6}(q)-\Theta_{5,6}(q))
                 \right\rangle\,.
\end{align}
Note the appearance of a $\log^2(q)$ term which indicates a more
involved indecomposable structure of the modules than a simple rank two
Jordan cell. Note also that a particular linear combination of otherwise
very benign Theta-functions shows up in ${\cal M}_3$ in order to make
this set close under modular transformations. This linear combination is
removed from set ${\cal M}_1$. As written down, these sets close under
the $S$-transformation $\tau\mapsto -\frac{1}{\tau}$. Under the 
$T$-transformation $\tau\mapsto\tau+1$, the sets ${\cal M}_1$ and ${\cal M}_2$
map into themselves, while ${\cal M}_3$ maps to ${\cal M}_3 \cup {\cal M}_1$.
The $S$-matrices for the first two sets are standard. For the third set,
it reads
\def\p{\phantom{-}}
\be\label{eq:S3}
  S^{(3)} = \left(\begin{array}{llll}
    \p 1 & \p 0    & \p 0 & \p 0              \\
    \p 0 & \p 0    & \p 0 & -\frac{1}{4\pi^2} \\
     -12 & \p 0    &   -1 & \p 0              \\
    \p 0 & -4\pi^2 & \p 0 & \p 0  
  \end{array}\right)\,,
\ee
which has determinant one. It comes as a surprise that the Eisenstein series
$E_2$ appears as solution to a modular differential equation. 

The corresponding conformal weights are as follows. The functions in
${\cal M}_1$ have asymptotics compatible with conformal weights $h\in
\{-1/24, 0, 1/8, 1/3, 5/8, 35/24\}$. The functions in ${\cal M}_2$ are
compatible with $h\in \{0, 1/8, 1/3, 5/8, 1\}$, and finally the functions
in ${\cal M}_3$ are all compatible with $h=0$. The simplest candiates for 
physical $h=0$ characters which could be compared to the work by Pearce
{\em el al.} \cite{Pearce01}, are given by the expression
\be
  \chi^{(j)}_{h=0}(q) = \frac{1}{\eta(q)}
  \sum_{n}(-1)^n q^{h(\frac{3}{2}n+\frac{1}{2}j)}
\ee
for $j=0,1,2$ and $h(k)=\frac{1}{3}k(2k-1)+\frac{1}{24}$. 
These functions agree with their
results up to the orders provided. But $j=0,1$ cannot yield the true characters
of the physical (rank one) $h=0$ representations. The reason is that they are 
not really linearly combined out of functions of all three sets ${\cal M}_i$. 
Namely,
\begin{align}
  \chi^{(0)}_{h=0}(q) &= \frac{1}{\eta(q)}\Big(\Theta_{1,6}(q)-
                         \frac{7}{12}\Theta_{5,6}(q)-
                         \frac{1}{12}(\partial\Theta)_{5,6}(q)\Big)\,,\\
  \chi^{(1)}_{h=0}(q) &= \frac{1}{\eta(q)}\Big(\Theta_{1,6}(q)-
                         \frac{5}{12}\Theta_{5,6}(q)+
                         \frac{1}{12}(\partial\Theta)_{5,6}(q)\Big)\,,\\
  \chi^{(j=2)}_{h=-1/3}(q) &= \frac{1}{6\eta(q)}\Big(
     3\Theta_{3,6}(q) + (\partial\Theta)_{3,6}(q)\Big)\,. 
\end{align}
only involve $\frac{1}{\eta}(\Theta_{1,6}-\Theta_{5,6})\equiv 1$ from ${\cal
M}_3$, which forms a trivial irreducible subrepresentation under the action of
the modular group, as can be read off (\ref{eq:S3}). 
However, the characters of the CFT should form a basis for a
representation of the modular group, in which this representation does not
decompose into smaller ones. This is only possible, if the physical $h=0$ 
representations have
characters which involve modular forms from all three sets ${\cal M}_i$ in
such a way that $E_2$ must be part of the linear combinations. This
requirement together with the condition, that the $q$-series must have 
non-negative integer coefficients greatly limits the possible linear 
combinations.

Of course, by now much more is known about logarithmic extensions of 
minimal models and especially the minimal model at $c=0$. See for example
\cite{Rasmussen:2008ii,Gaberdiel:2009ug,Gaberdiel:2010rg,Vasseur:2011ud} 
and references therein. In particular, the characters of all the 
representations of the logarithmically extended minimal model with $c=0$ have
been computed. It turns out that one needs the $h=0$ characters up to order
20 to uniquely fix them in terms of the modular functions from our sets
${\cal M}_i$. This is to be expected, as the modular differential equation has
order 20, which was first given in \cite{Flohr:unpublished,Adamovic:2009xs}. 
Using the notation from \cite{Rasmussen:2008ii,Gaberdiel:2009ug},
we find for the characters of the irreps
\begin{align}
	\chi_{ {\cal W}(0)} &= \frac{1}{\eta}\Big(
		\Theta_{1,6}-\Theta_{5,6}
		\Big) \equiv 1\,,\\
	\chi_{ {\cal W}(1)} &= \frac{1}{\eta}\Big(
		\frac{-11}{720}\Theta_{1,6}
		+ \frac{251}{720}\Theta_{5,6}
		+ \frac{10}{720}(\partial\Theta)_{1,6}
		+ \frac{70}{720}(\partial\Theta)_{5,6}
		\Big) + \frac{1}{720}E_2\,,\\
	\chi_{ {\cal W}(2)} &= \frac{1}{\eta}\Big(
		\frac{-11}{720}\Theta_{1,6}
		+ \frac{131}{720}\Theta_{5,6}
		+ \frac{10}{720}(\partial\Theta)_{1,6}
		- \frac{50}{720}(\partial\Theta)_{5,6}
		\Big) + \frac{1}{720}E_2\,,\\
	\chi_{ {\cal W}(5)} &= \frac{1}{\eta}\Big(
		\frac{71}{720}\Theta_{1,6}
		+ \frac{169}{720}\Theta_{5,6}
		- \frac{70}{720}(\partial\Theta)_{1,6}
		- \frac{10}{720}(\partial\Theta)_{5,6}
		\Big) - \frac{1}{720}E_2\,,\\
	\chi_{ {\cal W}(7)} &= \frac{1}{\eta}\Big(
		\frac{-49}{720}\Theta_{1,6}
		+ \frac{169}{720}\Theta_{5,6}
		+ \frac{50}{720}(\partial\Theta)_{1,6}
		- \frac{10}{720}(\partial\Theta)_{5,6}
		\Big) - \frac{1}{720}E_2\,,\\
	\chi_{ {\cal W}(1/3)} &= \frac{1}{\eta}\Big(
		\frac36\Theta_{3,6}+\frac16(\partial\Theta)_{3,6}
		\Big)\,,\\
	\chi_{ {\cal W}(10/3)} &= \frac{1}{\eta}\Big(
		\frac36\Theta_{3,6}-\frac16(\partial\Theta)_{3,6}
		\Big)\,,\\
	\chi_{ {\cal W}(1/8)} &= \frac{1}{\eta}\Big(
		\frac46\Theta_{2,6}+\frac16(\partial\Theta)_{2,6}
		\Big)\,,\\
	\chi_{ {\cal W}(5/8)} &= \frac{1}{\eta}\Big(
		\frac26\Theta_{4,6}+\frac16(\partial\Theta)_{4,6}
		\Big)\,,\\
	\chi_{ {\cal W}(21/8)} &= \frac{1}{\eta}\Big(
		\frac46\Theta_{4,6}-\frac16(\partial\Theta)_{4,6}
		\Big)\,,\\
	\chi_{ {\cal W}(33/8)} &= \frac{1}{\eta}\Big(
		\frac26\Theta_{2,6}-\frac16(\partial\Theta)_{2,6}
		\Big)\,,\\
	\chi_{ {\cal W}(-1/24)} &= \frac{1}{\eta}\Big(
		\Theta_{0,6}
		\Big)\,,\\
	\chi_{ {\cal W}(35/24)} &= \frac{1}{\eta}\Big(
		\Theta_{6,6}
		\Big)\,.
\end{align}
This shows, that $E_2$ is indeed part of all non-trivial irreps with integer
conformal weights. In particular, it contributes to the two non-trivial and
inequivalent physical $h=0$ rank one representations ${\cal W}$ and ${\cal Q}$, 
$\chi_{ {\cal W}} = \chi_{ {\cal W}(0)} + \chi_{ {\cal W}(1)}$ and
$\chi_{ {\cal Q}} = \chi_{ {\cal W}(0)} + \chi_{ {\cal W}(2)}$. These, as all
characters of the indecomposable representations, are linear combinations of the
characters of the irreps given above.

In conclusion, it is 
satisfying to see that the modular differential equation and some 
pre-knowledge on the vacuum character can tell as quite a few things about 
a conformal field theory, even in the notoriously difficult case $c=0$.


\section{Bosonic and Fermionic Expressions for Characters} \label{ref:chars-fermionic}

Characters can be written in a closed form in more than one way. In the following, we focus on the ``bosonic'' and on the ``fermionic'' character representations.
The character of a free, chiral boson with
momenta $p\in\Z_{\geq 1}$ is given by $\frac{1}{(q)_{\infty}}$. Truncations of the representation space through subsequent subtraction and addition of singular vectors will be encoded in
the numerator. Thus, character expressions obtained in this way have been termed bosonic.
In addition to the bosonic character expressions, there exists the concept of fermionic quasi-particle sum
representations for a character, also called fermionic expressions, which first
appeared under this name in \cite{KM93}. The fermionic expressions are interesting from both a
mathematical and a physical point of view and first occurred in an especially
simple form in the context of the Rogers-Schur-Ramanujan identities
\cite{Rog94,Sch17,RR19} (for $ a\in\{0,1\} $) 
\be \label{rsr}
\sum\limits_{n=0}^{\infty}\frac{q^{n(n+a)}}{(q)_n} = \prod\limits^{\infty}_{n=1}
\frac{1}{(1-q^{5n-1-a})(1-q^{5n-4+a})}
\ee
with the \emph{$q$-analogues}\index{$q$-analogue}
\be
(x;q)_n :=\prod_{i=0}^{n-1}(1-q^ix)\quad\text{and}\quad (q)_n:=(q;q)_n=\prod_{i=1}^{n}
(1-q^n)
\ee
of the Pochhammer symbol\index{Pochhammer symbol} and the classical factorial
function, respectively, and by definition
\be
(q)_0:=1\quad \text{and} \quad (q)_{\infty}:=\lim_{n\rightarrow\infty}(q)_n \ ,
\ee
the latter being the $q$-analogues of the classical gamma function. Note that
$(q)_{\infty}$ is up to factor of $q^{\frac{1}{24}}$ the modular form
$\eta(\tau)$ with $q=\e^{2\pi\ii\tau}$, the Dedekind $\eta$-function.
These identities coincide with the two characters of the minimal model
$\mathcal{M}(2,5)$ with central charge $c=-\frac{22}{5}$, which represents the
Yang-Lee model (up to an overall factor of $q^{\alpha}$ for some
$\alpha\in\C$). It is the smallest minimal model and contains only two primary
operators: the identity $\mathds{1}$ of dimension $(h,\bar h)=(0,0)$ and
another operator $\Phi$ of dimension $(-\frac{1}{5},-\frac{1}{5})$. By using
Jacobi's triple product identity \cite{Jac29}, defined for $z\neq 0$ and
$|q|<1$ (see \cite{And84}) as
\be \label{jacobi}
\sum_{n=-\infty}^{\infty}z^nq^{n^2}=\prod_{n=1}^{\infty}(1-q^{2n})(1+zq^{2n-1})(1+z^{-1}q^{2n-1})\ ,
\ee
the r.h.s. of \eqref{rsr} can be transformed to give a simple example of what
is called a \emph{bosonic-fermionic $q$-series
identity}\index{bosonic-fermionic $q$-series identity}:
\be \label{chars-bosonic}
\sum\limits_{n=0}^{\infty}\frac{q^{n(n+a)}}{(q)_n}=\frac{1}{(q)_{\infty}}\sum\limits_{n=-\infty}^{\infty}(q^{n(10n+1+2a)}-q^{(5n+2-a)(2n+1)})
\ee

For an instructive proof of the Jacobi triple product identity, see \cite{Kac96}, which employs a comparison of the characters computed from a fermionic
basis of the irreducible vacuum representation of charged free fermion system
with the character computed from a bosonic basis of the same representation
obtained by bosonization.

In general, it is always possible to write a minimal model character in a
product form and thus to obtain a Rogers-Ramanujan-type identity if $p=2s$ or
$p'=2r$, as has been demonstrated by Christe in \cite{Chr91}. To see
this, one employs the Jacobi triple product identity \eqref{jacobi} with the
replacements $q\mapsto q^{\frac{pp'}{2}}$ and $z\mapsto -q^{rs-\frac{pp'}{4}}$.
Product forms are also possible in the case $p=3s$ or $p'=3r$, but to show
this, the \emph{Watson identity}\index{Watson identity} \cite{Wat29} (see also
\cite[ex. 5.6]{GR90}) has to be used instead of the Jacobi identity. Christe
also proved in the same article that for other minimal model characters, no
product forms of this type exist.

The \emph{bosonic expressions}\index{bosonic character expressions} on the
r.h.s. of \eqref{chars-bosonic} correspond to two special cases of the general
character formula
for minimal models by Rocha-Caridi \cite{RoC84}.
Explictly, they are given by
\begin{align}
\chi^{5,2}_{1,1} & = 1+q^2+q^3+q^4+q^5+2q^6+2q^7+3q^8+3q^9+4q^{10}+4q^{11}\\
\nonumber &+6q^{12}+6q^{13}+8q^{14}+9q^{15}+11q^{16}+12q^{17}+15q^{18}+16q^{19}+20q^{20}+\ldots \\
\intertext{and}
\chi^{5,2}_{1,2} & = 1+q+q^2+q^3+2q^4+2q^5+3q^6+3q^7+4q^8+5q^9+6q^{10}+7q^{11}\\
\nonumber &+9q^{12}+10q^{13}+12q^{14}+14q^{15}+17q^{16}+19q^{17}+23q^{18}+26q^{19}+31q^{20}+\ldots \ .
\end{align}
Note that the coefficient of $q$ is zero because the vacuum is invariant under
$L_n$, $n\in\{ -1,0,1 \}$. Since the right hand side of \eqref{chars-bosonic}
is computed by eliminating null states from the state space of a free chiral
boson \cite{FF83}, it is referred to as bosonic form in the context of the Virasoro algebra.\footnote{See \cite{GL76} for an earlier account in terms of the character formula for affine Lie algebras.} Its signature is the
alternating sign, which reflects the subtraction of null vectors. The factor
$(q)_{\infty}$ keeps track of the free action of the Virasoro ``raising'' modes.
Furthermore, it can be expressed in terms of $\Theta$-functions (cf.
\ref{ref:theta}), which directly points out the modular transformation
properties of the character.

On the other hand, the left-hand side of \eqref{rsr} has a direct fermionic
quasi-particle interpretation for the states and hence it came to be known as fermionic sum
representation from some point on \cite{KM93}. In the first systematic study of fermionic expressions
\cite{KKMM93b}, sum representations for all characters of the unitary Virasoro
minimal models and certain non-unitary minimal models were given.
The list of expressions was augmented to all $p$ and $p'$ and certain $r$ and
$s$ in \cite{BMS98}.  Eventually, the fermionic expressions for the characters
of all minimal models were summarized in \cite{Wel05}.
Such a fermionic expression, which is a generalization of the left hand side of
\eqref{rsr}, is a linear combination of fundamental fermionic forms, as defined
in \eqref{fff}.

Fermionic character expressions in conformal field theory have various origins. Andrews and Gordon \cite{And74b} as well as Bressoud \cite{Bre80} found a family of generalizations of the Rogers-Ramanujan identities which Lepowsky and Wilson subsequently (as expressions closely related to \eqref{agid}) interpreted by means of twisted $Z$-algebras and twisted vertex operators \cite{LW81b,LW84,LW85}, through the implementation of the Pauli exclusion principle and generalizations thereof in terms of monomial bases of certain vector spaces. Lepowsky and Primc then delved into the special case of ``no twisting'' \cite{LP85}.\footnote{The same $Z$-algebras, both twisted and untwisted, were later rediscovered in \cite{FZ85,FZ87} in a variant form as ``parafermion algebras'', see also the last chapter of \cite{DL95b} for a precise ``dictionary'' between $Z$-algebras and parafermion algebras, in the untwisted setting. Another way in which the $Z$-algebras later came to be viewed is as the ``Abelian'' case of coset models, where the smaller structure is based on a Heisenberg Lie algebra. The $Z$-algebra program is now understood as an integral aspect of the representation theory of (generalized) vertex algebras, and it's now related to newer ideas like intertwining operators in vertex operator algebra theory. We thank James Lepowsky for pointing this out to us; see also \cite{Lep07}.} These were the first examples of what later came to be known -- in the context of the Virasoro algebra -- as bosonic and fermionic character expressions. Fermionic expressions are also known to arise from thermodynamic Bethe ansatz analysis
of integrable perturbations of conformal field theory \cite{KM90,KM92}
resulting in dilogarithm identities (cf. chapter \ref{ref:dilog}) which may be
lifted back \cite{Ter92} to fermionic expressions; furthermore they are known to arise from the scaling limit of
spin chains and $ADE$ generalizations of Lepowsky-Primc \cite{KM93,DKMM94} and
from spinon bases for WZW models \cite{BPS94,BLS95,BLS95b}. Some of these different
origins will be discussed in detail in the following sections.

\section{Nahm's Conjecture}\label{ref:nahm}

The question of how \emph{$q$-hypergeometric series}\index{$q$-hypergeometric
series} (i.e.~series of the form $\sum_{n=0}^{\infty}A_n(q)$ where $A_0(q)$ is
a rational function and $A_n(q)=R(q,q^n)A_{n-1}(q)\forall n\geq 1$ for some
rational function $R(x,y)$ with $\lim_{x\rightarrow 0}\lim_{y\rightarrow
0}R(x,y)=0$) are related to modular forms or modular functions
is an almost completely unsolved problem. But there is a conjecture by Werner
Nahm (see e.g.~\cite{Nah04}), which involves dilogarithms and torsion elements
of the \emph{Bloch group}\index{Bloch group} as well as rational conformal
field theories.
If $j=r$ in \eqref{fff}, then the fundamental fermionic form reduces (with a
rescaling $A\mapsto\frac{1}{2}A$)\footnote{This rescaling has only been done in
this section, since it makes the discussion of matrices $A$ and their inverses
easier. For the rest of this report, this rescaling is not necessary.} to the
$q$-hypergeometric series
\be
\label{nahm}f_{A,\vec{b},c}(\tau)=\sum_{\substack{\vec{m}\in(\mathbb{Z}_{\geq 0})^r \\ \text{restrictions}}}\frac{q^{\frac{1}{2}\vec{m}^tA\vec{m}+\vec{b}^t\vec{m}+c}}{(q)_{\vec{m}}} \ .
\ee
\emph{Nahm's conjecture}\index{Nahm's conjecture} has no complete answer to
this, but it makes a prediction which matrices $A\in M_r(\Q)$ can occur such
that \eqref{nahm} is a modular function, i.e.~whether there exist suitable
$\vec{b}\in\Q^r$ and $c\in\Q$ for a given matrix $A$. In particular, such a
function can only be modular when all solutions to a certain system of
algebraic equations depending on the coefficients of $A$, namely
\be \label{nahm-set_of_algs}
1-x_i=\prod_{j=1}^r x_j^{A_{ij}}  \Longleftrightarrow \sum_{j=1}^r A_{ij}\log (x_j) = \log (1-x_i)
\ee
(the same we will also encounter in chapter \ref{ref:dilog}), yield elements
$\sum_i [x_i]$ of finite order in the Bloch group of the algebraic numbers
\cite{Nah04}. 

The physical significance of this is that one expects that all the
$q$-hypergeometric series which are modular functions are characters of
rational conformal field theories. Given a matrix $A$, the modular forms for
the predicted possible combinations of vectors and constants span a
finite-dimensional vector space that is invariant under $\mathrm{PSL}(2,\Zop)$ for bosonic
CFTs (or under $\Gamma_0 (2)=\{\bsm a & b \\ c & d\esm\in \mathrm{PSL}(2,\Zop) \mid c\in
2\Z \}$ for fermionic CFTs), i.e.~the set of characters generated in this way
forms a finite-dimensional representation of the modular group, which is just
the definition of rationality of a conformal field theory.
Indeed, this is just what we will find in the subsequent analysis in this
report: The admissible matrices of rank one and two correspond to rational
theories, most of them to the minimal models.

In general, there exist fermionic expressions for all characters of the minimal
models. However, all but a finite number are not known to be of the type
\eqref{nahm}. Instead, they consist of finite linear combinations of
fundamental fermionic forms \eqref{fff}, i.e.~they involve finite $q$-binomial
coefficients. But nevertheless, it is usually possible to express all
characters of a given minimal model in terms of the same matrix $A$, albeit the
choice of the matrix for that given model is in general not unique. We will
comment more on that in the subsequent sections.

Note furthermore that the series of triplet $\mathcal{W}$-algebras, which are
logarithmic conformal field theories to be discussed later in this report, was
shown to be rational (in a broader sense to be defined later) with respect to
its extended $\mathcal{W}$-symmetry algebra. These theories are not rational
with respect to the Virasoro algebra alone as the symmetry algebra. By
presenting fermionic sum-representations of Nahm type \eqref{nahm} for the
characters of the whole series of $\mathcal{W}(2,2p-1,2p-1,2p-1)$ triplet
algebras ($p\geq 2$), thus leading to a new infinite set of bosonic-fermionic
$q$-series identities, we further support Nahm's conjecture and provide further
evidence that the triplet algebra series are well-defined new animals in the
zoo \cite{Flo03} of rational conformal field theories.

There are also fermionic expressions for characters of other theories than the
above mentioned, including for example the Kac-Peterson characters of the
affine Lie algebra $A_1^{(1)}$ \cite{KP84}.

A lot of matrices $A$, among them in particular one infinite series, have been
found for which Nahm's conjecture suggests that they should lead to modular
forms. However, a complete search has only been achieved for rank one and two
matrices, documented in \cite{Zag07}. Some of them are related to the \emph{Dynkin diagrams}\index{Dynkin
diagram} of the type $A$, $D$, $E$ or $T$, corresponding to the simple Lie
algebras.\footnote{Watch the notation problem: The matrix in the exponent of
the fermionic character expression is conventionally labeled $A$. This is not
to be confused with the $A$ series of Dynkin diagrams.} These diagrams have $r$ vertices if they are called $X_r$,
where $X$ is to be replaced by $A$, $D$, $E$ or $T$. In many cases, the matrix
$A$ in the quadratic form in \eqref{nahm} is just twice the inverse Cartan
matrix of a Dynkin diagram. On the other hand, it may also be half the Cartan
matrix itself. These two cases are to be regarded as the special cases of
another class, namely $A=C_{X_r}\otimes C_{Y_s}^{-1}$, where $X_r,Y_s\in\{
A,D,E,T \}$. Note that the \emph{Cartan matrix}\index{Cartan matrix} is in
one-to-one correspondence with a Dynkin diagram: For each vertex $i$ that is
connected to a vertex $j$ ($i,j\in\{ 1,\ldots , r\}$), set $A_{ij}A_{ji}$ equal
to the number of lines connecting these two vertices with the restriction that
$A_{ij},A_{ji}\in\Z_{\leq 1}$, set $A_{ii}=0\ \forall i\in \{ 1,\ldots , r\}$.
All the other entries are zero. An exception to this is $C_{T_r}$, which is
equal to $C_{A_r}$ in all components but in the lower right one:
$(C_{T_r})_{rr}=1$. $T_r$ is the \emph{tadpole graph}\index{tadpole graph}
corresponding to $A_{2r}$ folded in the middle such that vertices are pairwise
identified. Many $ADE$ related matrices of rank greater than two have also been
found to correspond to rational conformal field theories, especially the
inverse Cartan matrices. Examples of this kind will be discussed in section
\ref{ref:adet}. Modular forms with matrices of the second class can be found
e.g.~in Kac-Peterson characters later in this report, while the first class
is common to minimal models. But there are also other types, some of which
don't seem to fit in this pattern.

\section{$ADET$ Classification}\label{ref:adet}

The possibility of classifying fermionic character expressions according to simple Lie algebras is investigated further in this section.

All possible simple Lie algebras have been classified by Dynkin. Geometric constraints imply that there are only four infinite families and five exceptional cases. The infinite families are labeled by $A_n$, $B_n$, $C_n$ or $D_n$ and the exceptional cases by $G_2$, $F_4$, $E_6$, $E_7$ and $E_8$, where $n$ is the number of nodes of the corresponding \emph{Dynkin diagram}\index{Dynkin diagram}. Each of the above Lie algebras is assigned a Dynkin diagram, and the set of Dynkin diagrams is in one-to-one correspondence with the set of \emph{Cartan matrices}\index{Cartan matrix}.
If one labels the nodes of a Dynkin diagram by $a\in\{ 1,2,\ldots , n\}$, one can construct its Cartan matrix by setting $(C_{X_n})_{ab}(C_{X_n})_{ba}$ to the number of connecting lines between the nodes $a$ and $b$ of the Dynkin diagram to the Lie algebra $X_n$ and demanding that $(C_{X_n})_{ab}\leq 0$ and integer, and furthermore $(C_{X_n})_{aa}=2$.

The $ADE$ graphs play an important role in many places in mathematics and physics. In conformal field theory, for example, they can be used to classify modular invariant partition functions \cite{CIZ87} and, furthermore, the Cartan matrices also appear in the quadratic form in the exponent of the fermionic character expressions.

In the following, the conformal field theories whose fermionic character expressions correspond to the $ADE$ graphs are reported as well as the additional artificial series of \emph{tadpole graphs}\index{tadpole graph}, which also appear in fermionic expressions. The corresponding Dynkin diagrams can be found in e.g.~\cite{FS97}. When the matrix $A$ in the quadratic form is mentioned, the reader is always referred to \eqref{fff}.\footnote{Upon comparing \eqref{fff} with \eqref{nahm}, one notices that the exponents differ by a factor of $\frac{1}{2}$ in the definition of the matrix $A$. Strictly speaking, one should write $\frac{1}{2}A$ in the exponent of every fermionic form, since in general $A=C_{X_r}\otimes C_{Y_s}^{-1}$ for some $X_r,Y_s\in\{ A,D,E,T \}$ (as discussed in section \ref{ref:nahm}) and in most cases occurring in this report $X_r=A_1$ and thus $A=2C_{Y_s}^{-1}$. Therefore, since in most cases the factors $2$ and $\frac{1}{2}$ cancel, the general fundamental fermionic form \eqref{fff} is 
referred to in this report except in the single section \ref{ref:nahm} and where explicitly stated.}

\begin{description}
	\item[The $\mathbold{A_n}$ series]
corresponds to the unitary $\Z_{n+1}$ parafermionic theories with central charge $c_n=c_{\text{eff}}^n=\frac{2n}{n+3}$ \cite{FZ85,FL88}, where the effective central charge is defined by $c_\text{eff}=c-24h_\text{min}$. Lepowsky and Primc \cite{LW81b,LP85} found fermionic expressions with sum restrictions for the $\Z_{n+1}$ characters, the latter consisting of the $A_1$ string functions of level $k$ by Feingold and Lepowsky \cite{FL78}. Moreover, $A_1$ corresponds (trivially) to characters of the Ising model, namely
\begin{align} \label{ising-fer1}
\chi^{4,3}_{1,1} & = \sum_{\substack{ m=0 \\ m\equiv 0\pmod{2}}}^{\infty} \frac{q^{\frac{m^2}{2}-\frac{1}{48}}}{(q)_m}=\frac{\Th_{1,12}-\Th_{7,12}}{\eta} \\
\label{ising-fer2} \chi^{4,3}_{2,1} & = \sum_{\substack{ m=0 \\ m\equiv 1\pmod{2}}}^{\infty} \frac{q^{\frac{m^2}{2}-\frac{1}{48}}}{(q)_m}=\frac{\Th_{5,12}-\Th_{11,12}}{\eta}
\end{align}
and $A_2$ corresponds to the characters of $\mathcal{M}(6,5)$, namely
\be
\chi_{1,1}^{6,5}+\chi_{1,5}^{6,5}=\sum_{\substack{\vec{m}\vin{2} \\ m_1+2m_2\equiv 0\pmod{3}}}\frac{q^{\vec{m}^tC_{A_2}^{-1}\vec{m}-\frac{1}{30}}}{(q)_{\vec m}}
\ee
and
\be
\chi_{1,3}^{6,5}=\sum_{\substack{\vec{m}\vin{2} \\ m_1+2m_2\equiv a\pmod{3}}}\frac{q^{\vec{m}^tC_{A_2}^{-1}\vec{m}-\frac{1}{30}}}{(q)_{\vec m}}\ , \quad a\in\{-1,1\} \ .
\ee

Additionally, via $A=C_{A_n}\otimes C_{T_1}^{-1}=C_{A_n}$, this series also corresponds to the unitary series of minimal models via the unitary minimal model $\mathcal{M}(p+1,p)$ characters
\be \label{fe-unitary_mm}
q^{\alpha}\chi_{1,1}^{p+1,p}=\sum_{\substack{\vec{m}\vin{p-2} \\ m_i \text{ even}}}\frac{q^{\frac{1}{4}\vec{m}^tC_{A_{p-2}}\vec{m}}}{(q)_{m_1}}\prod_{i=2}^{p-2}\begin{bmatrix}((\mathds{1} - \frac{1}{2}C_{A_{p-2}})\vec{m})_i \\ m_i \end{bmatrix}_q \ .
\ee
This can be computed by the methods of Welsh \cite{Wel05} for all possible combinations of $r$ and $s$, but for simplicity, only the vacuum character is given here.
\item[The $\mathbold{D_n}$ series] corresponds to the unitary theory of a free boson compactified on a torus of radius $R=\sqrt{\frac{n}{2}}$ with central charge $c=c_{\text{eff}}=1$. This theory has characters $\frac{\Theta_{\lambda,k}}{\eta}$ for $\lambda\in\{-k+1, \ldots , k\}$ with $\lambda=0$ denoting the vacuum character. The fermionic expressions for these characters can be all be written with the inverse Cartan matrix of $D_n$ in the quadratic form. We will discuss this later on, when we derive the fermionic expressions for the $c_{p,1}$ series of logarithmic conformal field theories, where we will see that the whole $c_{p,1}$ series corresponds to the $D_n$ series, i.e.~the quadratic form in the fermionic character expressions is
\be
\vec m^t C_{D_p}^{-1}\vec m
\ee
for all characters of the $c_{p,1}$ model. The sum restrictions state that the sum $m_{n-1}+m_{n}$ has to be either even or odd, depending on the chosen character of the model. For the subset of characters that are of the form $\frac{\Theta_{\lambda,k}}{\eta}$, both restrictions admit a realization. Note furthermore that due to the coincidence $D_3=A_3$, some character functions corresponding to these two series are related.
\item[The exceptional algebra $\mathbold{E_6}$] corresponds to the unitary minimal model $\mathcal{M}(7,6)$ that is the \emph{tricritical three-state Potts model}\index{tricritical three-state Potts model} \cite{FZ87} with central charge $c=\frac{6}{7}$, namely
\be
\chi_{1,1}^{7,6}+\chi_{5,1}^{7,6}=\sum_{\substack{\vec{m}\vin{6} \\ m_1-m_2+m_4-m_5\equiv 0\pmod{3}}}\frac{q^{\vec{m}^tC_{E_6}^{-1}\vec{m}-\frac{c}{24}}}{(q)_{\vec m}}
\ee
and
\be
\chi_{3,1}^{7,6}=\sum_{\substack{\vec{m}\vin{6} \\ m_1-m_2+m_4-m_5\equiv a\pmod{3}}}\frac{q^{\vec{m}^tC_{E_6}^{-1}\vec{m}-\frac{c}{24}}}{(q)_{\vec m}} \ , \quad a\in\{-1,1\}
\ee
with
\be
C_{E_6}^{-1}=\begin{pmatrix}
\frac{4}{3}	& \frac{2}{3}	& 1	& \frac{4}{3}	& \frac{5}{3}	& 2	\\
\frac{2}{3}	& \frac{4}{3}	& 1	& \frac{5}{3}	& \frac{4}{3}	& 2	\\
1		& 1		& 2	& 2		& 2		& 3	\\
\frac{4}{3}	& \frac{5}{3}	& 2	& \frac{10}{3}	& \frac{8}{3}	& 4	\\
\frac{5}{3}	& \frac{4}{3}	& 2	& \frac{8}{3}	& \frac{10}{3}	& 4	\\
2		& 2		& 3	& 4		& 4		& 6
\end{pmatrix}\quad .
\ee
By adding a suitable vector $\vec b$ to the exponent in the fermionic expression or by changing the sum restrictions, the other characters of $\mathcal{M}(7,6)$ might also be found to have fermionic representations of this type, but so far, none are known.
\item[The exceptional algebra $\mathbold{E_7}$] corresponds to the tricritical Ising unitary model $\mathcal{M}(5,4)$ with central charge $c=\frac{7}{10}$ mentioned in the previous section.
Here,
\be
\chi_{1,1}^{5,4} = \sum_{\substack{\vec{m}\vin{7} \\ m_1+m_3+m_6\equiv 0\pmod{2}}}\frac{q^{\vec{m}^tC_{E_7}^{-1}\vec{m}-\frac{c}{24}}}{(q)_{\vec m}}
\ee
and
\be
\chi_{3,1}^{5,4} = \sum_{\substack{\vec{m}\vin{7} \\ m_1+m_3+m_6\equiv 1\pmod{2}}}\frac{q^{\vec{m}^tC_{E_7}^{-1}\vec{m}-\frac{c}{24}}}{(q)_{\vec m}} \ .
\ee
\item[The exceptional algebra $\mathbold{E_8}$] corresponds to the Ising model $\mathcal{M}(4,3)$ and, as also mentioned in the previous section,
\be
\chi^{4,3}_{1,1} = \sum_{\vec{m}\vin{8}}\frac{q^{\vec{m}^tC_{E_8}^{-1}\vec{m}-\frac{1}{48}}}{(q)_{\vec m}} \ .
\ee
\item[The $\mathbold{T_n}$ series], often called $\frac{A_{2n}}{\Z_2}$, corresponds to the series of non-unitary Virasoro minimal models $\mathcal{M}(2n+3,2)$ with effective central charge $c_{\text{eff}}^{k}=\frac{2n}{2n+3}$. Their characters admit a product form \cite{ABS90,FNO92}, which is one side of the Andrews-Gordon identities \cite{Gor61,And74b,Bre80,And84}
\begin{align}\label{agid}
\sum_{m_1,\ldots,m_n=0}^{\infty}\frac{q^{M_1^2+\ldots+M_n^2+M_a+\ldots+M_n}}{(q)_{m_1}\cdots(q)_{m_n}}
=\prod_{\substack{m \nequiv 0 \pmod{2n+3} \\ m \nequiv \pm a \pmod{2n+3}}}(1-q^m)^{-1}
\end{align}
with $M_k:=m_1+\ldots+m_k$. Gordon gave the combinatorial and Andrews the analytical proof. The other side consists of the fermionic sum representations for the characters of $\mathcal{M}(2n+3,2)$. This original formulation can be rewritten in order to match the fermionic forms as
\be
\chi_{1,j}^{2n+3,2}(q)=q^{h_{1,j}^{2n+3,2}-\frac{c_{2n+3,2}}{24}}\sum_{\vec m\vin{n}}\frac{q^{\vec{m}^t C_{T_n}^{-1}\vec m + \vec{b}_{T_n}^{t}\vec{m}}}{(q)_{\vec m}}
\ee
with $C_{T_n}$ being the Cartan matrix of the tadpole graph which differs from $C_{A_n}$ only by a $1$ instead of a $2$ in the component $\left( C_{T_n} \right)_{nn}$ and $\vec{b}_{T_n}^{t}=(\underbrace{0,\ldots,0}_{n \text{ times}},1,2,\ldots,k-n)$. Note that the Andrews-Gordon identities reduce to the Rogers-Ramanujan identities for $n=1$ and $a\in\{1,2\}$, i.e.~for $\mathcal{M}(5,2)$.

\end{description}

Most of these expressions were found and verified by using Mathematica and explicit proofs were lacking for most of them \cite{KKMM93a}. But that situation changed during the following years. A particular example is the fermionic character expression for $\chi^{4,3}_{1,1}$ related to $E_8$, for which Warnaar and Pearce found a proof based on the dilute $A_3$ model \cite{WP94}. A different direction that allowed for many of the identities to be proven was found by Melzer \cite{Mel94a}. He observed that Virasoro characters have a natural \emph{finitized}\index{finitized characters} version in terms of \emph{path spaces}\index{path spaces} or \emph{corner-transfer matrix sums}\index{corner-transfer matrix} in the \emph{rough solid-on-solid (RSOS) model}\index{RSOS model} \cite{ABF84}. This method of proving the identities has been extended in \cite{Ber94} and references therein.
The fact that there are different fermionic expressions for a single character (in the sense that the matrix $A$ in the quadratic form is different) is demonstrated impressively by the vacuum character $\chi_{1,1}^{4,3}$ of the Ising model. There is a sum representation related to $A_1$ and a sum representation related to $E_8$. Let us discuss this. In \cite{KM90}, Klassen and Melzer investigated integrable massive scattering theories. There, the $ADE$ algebras describe certain perturbations of coset conformal field theories \cite{GKO85} related to $ADE$. These algebras are the same. For example, the energy perturbation of the Ising model, which is called \emph{Ising field theory}\index{Ising field theory}, corresponds to $A_1$ and to the conformal limit of Kaufman's representation of the general Ising model in the absence of a magnetic field in terms of a single, free fermion \cite{Kau49}, while the magnetic perturbation corresponds to a scattering theory of eight different particle species \cite{Zam89}. 
Later on in this report, when we demonstrate the quasi-particle interpretation of the fermionic character expressions, we will see that the $E_8$ character corresponds also to a system of eight quasi-particle species with exactly the charges in \cite{Zam89} reproduced by the sum restrictions. This is another example that different fermionic expressions for the same character point to different integrable perturbations of the conformal field theory in consideration.

Furthermore, symmetries of the character $\chi_{r,s}^{p,p'}$ with respect to its parameters
add to the non-uniqueness of a fermionic character expression. For instance,
\be
\chi_{\alpha r,s}^{p,\alpha p'}=\chi_{r,\alpha s}^{\alpha p,p'} \quad \alpha\in\Z_{\geq 1}\ , \ \langle p,\alpha p'\rangle = \langle \alpha p,p'\rangle = 1
\ee
implies that the characters of $\mathcal{M}(6,5)$ are related to those of $\mathcal{M}(10,3)$.

\section{Characters of the Triplet Algebras $\mathcal{W}(2,2p-1,2p-1,2p-1)$}

\subsection{Characters in Bosonic Form}

The triplet $\mathcal{W}$ algebras are rational conformal field theories \cite{GK96b,Flo96,CF06}, i.e.~the number of highest weight representations of the $\mathcal{W}$-algebra is finite, and the generalized character functions span a finite-dimensional representation of the modular group. They are rational in the generalized sense discussed in section \ref{ref:triplet}, since indecomposable representations occur. Knowing the vacuum character is sufficient in proving rationality of the theory.

One can calculate the $\mathcal{W}$ character of the vacuum representation by summing up all the Virasoro characters of the highest weight representations corresponding to integer values of $h$, the latter being given by
\be
h_{2k+1,1}=k^2p+kp-k \ .
\ee
All the corresponding primary fields belong to degenerate conformal families. By means of a standard free-field construction \cite{BPZ84,DF84,DF85a,DF85b}, it turns out that the representations with these highest weights $h_{2k+1,1}$ correspond to a set of relatively local chiral vertex operators $\Phi_{2k+1,1}$. It follows that the local chiral algebra can be extended by them. The conditions for the existence of well-defined chiral vertex operators \cite{Kau95,Kau00} result in abstract fusion rules which imply that the local chiral algebra generated by only the stress-energy tensor and the field $\Phi_{3,1}$ closes. Repeated application of the \emph{screening charge operator}\index{screening charge} $Q$ \cite{Fel89} on $\Phi_{3,1}$ generates a multiplet structure. Thus, one also has to take care of the $\mathfrak{su}(2)$ symmetry of the triplet of fields, which results in the multiplicity of the Virasoro representation $\ket{h_{2k+1,1}}$ being $2k+1$. E.g., since $h_{3,1}=2p-1$ and its multiplicity is three,
 it matches the fact that we have a triplet of fields in the algebra $\mathcal{W}(2,2p-1,2p-1,2p-1)$.
The vacuum representation of the ${\cal W}$-algebra can then be written as the following decomposition of the state space:
\be
{\cal H}_{\ket{0}}=\bigoplus\limits_{k \in \mathbb Z_{\geq 0}}(2k+1){\cal H}^{\text{Vir}}_{\ket{h_{2k+1,1}}}\ .
\ee

The embedding structure of Feigin and Fuks \cite{FF83} in the case of $p'=1$ implies that the Virasoro characters corresponding to $h_{2k+1,1}$ -- these are the only integer-valued weights for all $p$ -- are given by
\be
\chi_{2k+1,1}^{\text{Vir}}=\frac{q^{\frac{1-c_{p,1}}{24}}}{\eta(q)}\l q^{h_{2k+1,1}}-q^{h_{2k+1,-1}}\r \ .
\ee
It is thus possible to compute the vacuum character as
\begin{align}
\nonumber\chi_0^\mathcal{W}(q) & = \sum_{k\in\Z_{\geq 0}}(2k+1)\chi_{2k+1,1}^{\text{Vir}}(q)\nn\\
\nn&= \frac{q^{\frac{1-c_{p,1}}{24}}}{\eta(q)}\l \sum_{k\in\Z_{\geq 0}}(2k+1)q^{h_{2k+1,1}}-\sum_{k\in\Z_{\geq 0}}(2k+1)q^{h_{2k+1,-1}} \r \\
			& = \frac{q^{\frac{(p-1)^2}{4p}}}{\eta(q)}\l \sum_{k\in\Z_{\geq 0}}(2k+1)q^{h_{2k+1,1}}+\sum_{k\in\Z_{\leq 1}}(2k+1)q^{h_{2k+1,1}} \r \\
\notag&= \frac{q^{\frac{(p-1)^2}{4p}}}{\eta(q)}\l \sum_{k\in\Z}(2k+1)q^{h_{2k+1,1}} \r = \frac{1}{\eta(q)}\l \sum_{k\in\Z}(2k+1)q^{pk^2+kp-k+\frac{(p-1)^2}{4p}} \r \\
\nonumber			& = \frac{1}{p\eta(q)}\l \sum_{k\in\Z}(2pk+p)q^{\frac{(2pk+(p-1))^2}{4p}} \r = \frac{1}{p\eta(q)}\l (\partial\Theta)_{p-1,p}(q)+\Theta_{p-1,1}(q) \r \ ,
\end{align}
where the symmetry property $h_{r,s}=h_{-r,-s}$ has been used and the $\Theta$-functions as defined in \ref{ref:theta}.
The $h$-values of a given $\mathcal{W}$-algebra can be calculated by use of the free-field construction, using Jacobi identities and null field constraints (cf.~section \ref{ref:triplet}). The corresponding characters may be calculated as follows: The modular differential equation (see e.g.~\cite{Flo96}) may be used to compute as many terms of the $q$-expansion of the character as are necessary to unambigiously identify the corresponding function, because the requirement of that function to be a modular form implies strong restrictions on that function.
It turns out that if we assume that $c_{3p,3}=c_{p,1}$ corresponds to a minimal model, which of course it doesn't since $3p$ and $3$ are not coprime, it is possible to read the resulting $h$-values of the given $c_{p,1}$ theory off that enlarged Kac table.

$\frac{\Theta_{\lambda,k}(\tau)}{\eta(\tau)}$ is a modular form of weight zero with respect to the generators ${\cal T}:\ \tau \mapsto \tau +1$ and ${\cal S}:\ \tau \mapsto -\frac{1}{\tau}$ of the modular group $\mathrm{PSL}(2,\Zop)$. But since $\frac{(\partial\Theta)_{\lambda,k}(\tau)}{\eta(\tau)}$ is a modular form of weight one with respect to ${\cal S}$ (cf. section \ref{ref:theta}), some of the above character functions are of inhomogeneous modular weight, thus leading to $S$-matrices with $\tau$-dependent coefficients. However, adding
\be
(\nabla \Theta )_{\lambda,k}(\tau) = \frac{\log q}{2 \pi \ii}\sum_{n\in \mathbb Z}(2kn+\lambda)q^{\frac{(2kn+\lambda)^2}{4k}}\ ,
\ee
one finds a closed finite-dimensional representation of the modular group with constant $S$-matrix coefficients.

After all, it turns out that a complete set of character functions for the $c_{p,1}$ models that is closed under modular transformations \cite{Flo97} is given by:
\begin{subequations}\label{tripletw-chars}
\begin{align}
\label{triplet-char-0}\chi_{0,p} & = \frac{\Theta_{0,p}}{\eta} 
	& \text{\textrm{of representation}}\qquad 
	& {\cal W}({\textstyle h_{1,p}^{p,1}}) \\
\chi_{p,p} & = \frac{\Theta_{p,p}}{\eta} 
	& 
	& {\cal W}({\textstyle h_{1,2p}^{p,1}}) \\
\chi_{\lambda,p}^{+} & = \frac{(p-\lambda)\Theta_{\lambda,p}+(\partial\Theta)_{\lambda,p}}{p\eta} 
	& 
	& {\cal W}({\textstyle h_{1,p-\lambda}^{p,1}}) \\
\chi_{\lambda,p}^{-} & = \frac{\lambda\Theta_{\lambda,p}-(\partial\Theta)_{\lambda,p}}{p\eta} 
	& 
	& {\cal W}({\textstyle h_{1,3p-\lambda}^{p,1}})\\
\label{tilde_char_plus}\tilde{\chi}_{\lambda,p}^{+} & = \frac{\Theta_{\lambda,p}+\ii\alpha\lambda(\nabla\Theta)_{\lambda,p}}{\eta} 
	& 
	& {\cal R}({\textstyle h_{1,p+\lambda}^{p,1}}) \\
\label{tilde_char_minus}\tilde{\chi}_{\lambda,p}^{-} & = \frac{\Theta_{\lambda,p}-\ii\alpha(p-\lambda)(\nabla\Theta)_{\lambda,p}}{\eta} 
	& 
	& {\cal R}({\textstyle h_{1,p+\lambda}^{p,1})}
\end{align}\end{subequations}
where $0<\lambda<k$, $k=pp'=p$, $\lambda=pr-p's=pr-s$ and with the \emph{Jacobi-Riemann $\Theta$-function} and the \emph{affine $\Theta$-function} defined as in \ref{ref:theta}.

Note that \eqref{tilde_char_plus} and \eqref{tilde_char_minus} are not characters of representations in the usual sense. Actually, these are regularized character functions and the $\alpha$-dependent part has an interpretation as torus vacuum amplitudes \cite{FG06}. In the limit $\alpha\rightarrow 0$, they become the characters of the full reducible but indecomposable representations.

\subsection{Fermionic Character Expressions for $\mathcal{W}(2,3,3,3)$}
\label{ref:w2333-fer}

Fermionic sum representation for the $c_{p,1}$ models have been presented in \cite{FGK07} and proven in \cite{War07}. In this section, the derivation of the fermionic formulae for the case of $p=2$ is reviewed, while the case of $p>2$ is postponed to the next section.

In the case of $p=2$, the bosonic characters read:
\begin{subequations}\label{w2333-chars-bos}
\begin{align}
\label{h11}\chi_{1,2}^{+} & = \frac{\Theta_{1,2}+(\partial\Theta)_{1,2}}{2\eta} 
& {\textstyle \text{\normalsize \textrm{vacuum irrep }} \mathcal{W}(0)\text{ \normalsize \textrm{to} } h_{1,1}=0} \\
\label{h12}\chi_{0,2} & = \frac{\Theta_{0,2}}{\eta} & {\textstyle \text{\normalsize \textrm{irrep to} } h_{1,2}=-\frac{1}{8}}\\
\label{tilde_char_simplified}\chi_{1,2} & = \frac{\Theta_{1,2}}{\eta} 
& {\textstyle \text{\textrm{\normalsize indecomp. rep }} \mathcal{R}(0) \supset \mathcal{W}(0) \text{ \normalsize \textrm{to} } h_{1,3}=0 } \\
\label{h14}\chi_{2,2} & = \frac{\Theta_{2,2}}{\eta} & {\textstyle \text{\textrm{\normalsize irrep to} } h_{1,4}=\frac{3}{8}}\\
\label{h15}\chi_{1,2}^{-} & = \frac{\Theta_{1,2}-(\partial\Theta)_{1,2}}{2\eta} & {\textstyle \text{\textrm{\normalsize irrep to} } h_{1,5}=1} & .
\end{align}\end{subequations}
When $\alpha \rightarrow 0$, the general forms \eqref{tilde_char_plus} and \eqref{tilde_char_minus} lead to the character expression \eqref{tilde_char_simplified} \cite{Kau95,Flo97}. Actually, there exist two indecomposable representations, $\mathcal{R}_0$ and $\mathcal{R}_1$ (cf. section \ref{ref:triplet}), which, however, share the same character.


In the following, the fermionic expressions for $\frac{\Theta_{\lambda,2}(\tau)}{\eta(\tau)}$, $0\leq\lambda\leq 2$, are being calculated at first. In this case, the bosonic expressions can be straightforward transformed to the fermionic ones: At first,
\begin{align}\label{w2333-a}
\frac{\Theta_{\lambda,k}}{(q)_\infty} & = \sum_{n=-\infty}^{+\infty}\frac{q^{\frac{(2kn+\lambda)^2}{4k}}}{(q)_\infty} \\
& = \frac{1}{(q)_\infty}\left( q^{\frac{\lambda^2}{4k}}+\sum_{n=1}^\infty q^{\frac{(2kn-\lambda)^2}{4k}}+\sum_{n=1}^\infty q^{\frac{(2kn+\lambda)^2}{4k}} \right) \ .\nonumber
\end{align}
Then, an identity
\be \label{durfee}
\sum_{n=0}^{\infty}\frac{q^{n^2+nk}}{(q)_n(q)_{n+k}}=\frac{1}{(q)_{\infty}}
\ee
that can be proven using Durfee squares or the $q$-analogue of Kummer's theorem (see e.g.~\cite[pp. 21,28]{And84})
is employed to turn \eqref{w2333-a} into
\be
\sum_{m=0}^\infty\frac{q^{m^2}}{(q)_m^2}+\sum_{n_1=1}^\infty \sum_{m_1=0}^\infty\frac{q^{m_1^2+m_1(2n_1)+\frac{(k(2n_1)-\lambda)^2}{4k}}}{(q)_{m_1}(q)_{m_1+2n_1}}+\sum_{n_2=1}^\infty \sum_{m_2=0}^\infty\frac{q^{m_2^2+m_2(2n_2)+\frac{(k(2n_2)+\lambda)^2}{4k}}}{(q)_{m_1}(q)_{m_1+2n_2}} \ .
\ee
Setting $n_1=\frac{m_2-m_1}{2}$ and $n_2=\frac{m_1-m_2}{2}$ leads to
\be \label{w2333-b}
\sum_{m=0}^\infty\frac{q^{m^2}}{(q)_m^2}+\sum_{\substack{0\leq m_1 < m_2=0 \\m_1+m_2\equiv 0 \pmod{2}}}^\infty\sum_{\substack{0\leq m_2 < m_1=0 \\m_1+m_2\equiv 0 \pmod{2}}}^\infty\frac{q^{\frac{k}{4}(m_1^2+m_2^2)+\frac{2-k}{2}m_1m_2+\frac{\lambda}{2}(m_1-m_2)+\frac{\lambda^2}{4k}}}{(q)_{m_1}(q)_{m_2}} \ .
\ee
On the other hand,
\begin{alignat}{2}
\nonumber\frac{\Theta_{\lambda,k}}{(q)_\infty}  = \sum_{n=-\infty}^{+\infty}\frac{q^{\frac{(2kn+\lambda)^2}{4k}}}{(q)_\infty} = &\sum_{n_1=1}^\infty \sum_{m_1=0}^\infty\frac{q^{m_1^2+m_1(2n_1-1)+\frac{-2k(k-\lambda)(2n_1-1)+(k-\lambda)^2+k^2(2n_1-1)^2}{4k}}}{(q)_{m_1}(q)_{m_1+2n_1-1}} \\ 
& +\sum_{n_2=1}^\infty \sum_{m_2=0}^\infty\frac{q^{m_2^2+m_2(2n_2-1)+\frac{2k(k-\lambda)(2n_2-1)+(k-\lambda)^2+k^2(2n_2-1)^2}{4k}}}{(q)_{m_2}(q)_{m_2+2n_2-1}} \ .
\end{alignat}
Setting $n_1=\frac{m_2-m_1+1}{2}$ and $n_2=\frac{m_1-m_2+1}{2}$ implies
\be \label{w2333-c}
\sum_{\substack{0\leq m_1 < m_2=0 \\m_1+m_2\equiv 1 \pmod{2}}}^\infty\sum_{\substack{0\leq m_2 < m_1=0 \\m_1+m_2\equiv 1 \pmod{2}}}^\infty\frac{q^{\frac{k}{4}(m_1^2+m_2^2)+\frac{2-k}{2}m_1m_2+\frac{k-\lambda}{2}(m_1-m_2)+\frac{(k-\lambda)^2}{4k}}}{(q)_{m_1}(q)_{m_2}} \ .
\ee
Thus, from \eqref{w2333-b} and \eqref{w2333-c},
\begin{subequations}\label{fe-theta_by_eta}
\begin{align}
\Lambda_{\lambda,k}(\tau) =  \frac{\Theta_{\lambda,k}(\tau)}{\eta(\tau)} \label{fe-theta_by_eta1} & = \sum_{\substack{\vec{m}=0 \\ m_1+m_2\equiv 0  \pmod{2}}}^{\infty}\frac{q^{\frac{1}{4}\vec{m}^t\bsm k & 2-k\\2-k & k \esm\vec{m}+\frac{1}{2}\bsm \psg \lambda\\-\lambda \esm^{\!{\!t}} \vec{m}+\frac{\lambda^2}{4k}-\frac{1}{24}}}{(q)_{\vec{m}}} \\
\label{fe-theta_by_eta2} & = \sum_{\substack{\vec{m}=0 \\ m_1+m_2\equiv 1 \pmod{2}}}^{\infty}\frac{q^{\frac{1}{4}\vec{m}^t\bsm k & 2-k\\2-k & k \esm \vec{m}+\frac{1}{2}\bsm -(k-\lambda)\\ \psg k-\lambda \esm^{\!{\!t}} \vec{m}+\frac{(k-\lambda)^2}{4k}-\frac{1}{24}}}{(q)_{\vec{m}}} \ .
\end{align}\end{subequations}
This two-fold $q$-hypergeometric series has been given without explicit proof in \cite{KMM93}. (Note that \eqref{fe-theta_by_eta} is not unique just as \eqref{theta}: According to \eqref{theta_symmetries} the vector may be changed in certain ways along with the constant.)
These are fermionic expressions for \eqref{h12} to \eqref{h14}. We obtain the fermionic expressions of the remaining two characters as follows \cite{FGK07}:
Note that $\frac{(\partial\Theta)_{1,2}}{\eta^3(q)}=1$ and hence
\be
\chi_{1,2}^{\pm} = \frac{\Theta_{1,2}}{2\eta}\pm \frac{1}{2}\eta^2 \ .
\ee
We then use an identity
\be
\eta(q)=q^{\frac{1}{24}}\sum_{n=0}^{\infty}\frac{(-1)^nq^{{n+1\choose 2}}}{(q)_n}
\ee
by Euler (cf. \cite{Zag07} for a simple proof). This identity may be squared, leading to
\be
\label{fe-eta_squared}\eta^2(q)=\tilde{\eta}^2(q,-1) \quad \text{with} \quad \tilde{\eta}^2(q,z)=\sum_{\vec{m}=0}^{\infty}\frac{q^{\frac{1}{2}\vec{m}^t\bsm 1 & 0\\0 & 1 \esm \vec{m}+\frac{1}{2}\bsm 1\\1 \esm^{\!{\!t}} \vec{m}+\frac{1}{12}}z^{m_1+m_2}}{(q)_{\vec{m}}} \ .
\ee
It is possible to transform the fermionic expression of $\chi_{1,2}$ which was obtained in \eqref{fe-theta_by_eta} into
\begin{align}
&\notag\sum_{\substack{\vec{m}=0 \\ m_1+m_2\equiv 0 \pmod{2}}}^{\infty}\frac{q^{\frac{1}{2}\vec{m}^t\bsm 1 & 0\\0 & 1 \esm\vec{m}+\frac{1}{2}\bsm \psg 1\\-1 \esm^{\trans} \vec{m}}}{(q)_{\vec{m}}}\\
= &\sum_{\substack{\vec{m}=0 \\ m_1+m_2\equiv 0 \pmod{2}}}^{\infty}\frac{q^{\frac{1}{2}\vec{m}^t\bsm 1 & 0\\0 & 1 \esm\vec{m}+\frac{1}{2}\bsm \psg 1\\-1 \esm^{\trans} \vec{m}}}{2(q)_{\vec{m}}}+\sum_{\substack{\vec{m}=0 \\ m_1+m_2\equiv 1 \pmod{2}}}^{\infty}\frac{q^{\frac{1}{2}\vec{m}^t\bsm 1 & 0\\0 & 1 \esm\vec{m}+\frac{1}{2}\bsm \psg 1\\-1 \esm^{\trans} \vec{m}}}{2(q)_{\vec{m}}} \nonumber\\
=&\biggl(\sum_{m_1=0}^{\infty}\frac{q^{\frac{m_1^2-m_1}{2}}}{(q)_{m_1}}\biggr) \biggl(\sum_{\substack{m_2=0 \\ m_2\equiv m_1 \pmod{2}}}^{\infty}\frac{q^{\frac{m_2^2+m_2}{2}}}{(q)_{m_2}}+\sum_{\substack{m_2=0 \\ m_2\equiv m_1+1 \pmod{2}}}^{\infty}\frac{q^{\frac{m_2^2+m_2}{2}}}{(q)_{m_2}}\biggr) \ .
\end{align}
By using
\begin{align}
\sum_{m=0}^{\infty}\frac{q^{\frac{m_2^2+m_2}{2}}}{(q)_m} = \frac{1}{2}\sum_{m=0}^{\infty}\frac{q^{\frac{m_2^2+m_2}{2}}}{(q)_m} \ ,
\end{align}
which holds because
\begin{align}
\nonumber \sum_{m=0}^{\infty}\frac{q^{\frac{m_2^2+m_2}{2}}}{(q)_m}  = & \sum_{m=0}^{\infty}\frac{q^{\frac{m_2^2+m_2}{2}}(1-q^{m+1})}{(q)_{m+1}}\\
= &\sum_{m=0}^{\infty}\frac{q^{\frac{m_2^2+m_2}{2}}}{(q)_{m+1}}-\sum_{m=0}^{\infty}\frac{q^{\frac{m_2^2+m_2}{2}+m+1}}{(q)_{m+1}} = \sum_{m=1}^{\infty}\frac{q^{\frac{m_2^2-m_2}{2}}}{(q)_m}-\sum_{m=1}^{\infty}\frac{q^{\frac{m_2^2+m_2}{2}}}{(q)_{m+1}} \ ,
\end{align}
$\chi_{1,2}$ may be written as
\begin{align}
q^{\frac{1}{24}}\chi_{1,2} = \sum_{\vec{m}=0}^{\infty}\frac{q^{\frac{1}{2}\vec{m}^t\bsm 1 & 0\\0 & 1 \esm\vec{m}+\frac{1}{2}\bsm 1\\1 \esm^{\trans} \vec{m}}}{(q)_{\vec{m}}} \ ,
\end{align}
finally leading to
\begin{alignat}{2}
\nonumber q^{\frac{1}{24}}\chi_{1,2}^{\pm} & = & & \frac{\Theta_{1,2}}{2\eta}\pm \frac{\eta^2}{2} \\
\nonumber & = & & \frac{1}{2}\sum_{\vec{m}=0}^{\infty}\frac{q^{\frac{1}{2}\vec{m}^t\bsm 1 & 0\\0 & 1 \esm\vec{m}+\frac{1}{2}\bsm \psg 1\\1 \esm^{\trans} \vec{m}}}{(q)_{\vec{m}}} \pm \frac{1}{2}\sum_{\vec{m}=0}^{\infty}\frac{q^{\frac{1}{2}\vec{m}^t\bsm 1 & 0\\0 & 1 \esm \vec{m}+\frac{1}{2}\bsm 1\\1 \esm^{\!{\!t}} \vec{m}+\frac{1}{12}}(-1)^{m_1+m_2}}{(q)_{\vec{m}}} \\
& = & & \sum_{\substack{\vec{m}=0 \\ m_1+m_2\equiv a \pmod{2}}}^{\infty}\frac{q^{\frac{1}{2}\vec{m}^t\bsm 1 & 0\\0 & 1 \esm\vec{m}+\frac{1}{2}\bsm 1\\1 \esm^{\trans} \vec{m}}}{(q)_{\vec{m}}}
\end{alignat}
with $a=0$ if the plus sign is chosen and $a=1$ if the minus sign is chosen.

Thus, also the remaining two characters yield expressions which consist of only one fundamental fermionic form. 

The following is a list of the fermionic expressions for all five characters of the logarithmic conformal field theory model corresponding to central charge $c_{2,1}=-2$ \cite{FGK07}:
\begin{subequations}
\begin{align}
\label{fe-h11}\chi_{1,2}^{+} & = \sum_{\substack{\vec{m}\vin{2} \\ m_1+m_2\equiv 0 \pmod{2}}}\frac{q^{\frac{1}{2}\vec{m}^t\bsm 1 & 0\\0 & 1 \esm\vec{m}+\frac{1}{2}\bsm 1\\ 1 \esm^{\!{\!t}}\vec{m}+\frac{1}{12}}}{(q)_{\vec{m}}} \\
\label{fe-h12}\chi_{0,2} & = \sum_{\substack{\vec{m}\vin{2} \\ m_1+m_2  \equiv 0 \pmod{2}}}\frac{q^{\frac{1}{2}\vec{m}^t\bsm 1 & 0\\0 & 1 \esm\vec{m}-\frac{1}{24}}}{(q)_{\vec{m}}} \\
\label{fe-h13}\chi_{1,2} & = \sum_{\substack{\vec{m}\vin{2} \\ m_1+m_2\equiv 0 \pmod{2}}}\frac{q^{\frac{1}{2}\vec{m}^t\bsm 1 & 0\\0 & 1 \esm\vec{m}+\frac{1}{2}\bsm \psg 1 \\-1 \esm^{\!{\!t}}\vec{m}+\frac{1}{12}}}{(q)_{\vec{m}}} \\
\label{fe-h14}\chi_{2,2} & = \sum_{\substack{\vec{m}\vin{2} \\ m_1+m_2\equiv 0 \pmod{2}}}\frac{q^{\frac{1}{2}\vec{m}^t\bsm 1 & 0\\0 & 1 \esm\vec{m}+\bsm \psg 1\\-1\esm^{\!{\!t}}\vec{m}+\frac{11}{24}}}{(q)_{\vec{m}}} \\
\label{fe-h15}\chi_{1,2}^{-} & = \sum_{\substack{\vec{m}\vin{2} \\ m_1+m_2\equiv 1 \pmod{2}}}\frac{q^{\frac{1}{2}\vec{m}^t\bsm 1 & 0\\0 & 1 \esm\vec{m}+\frac{1}{2}\bsm 1 \\1 \esm^{\!{\!t}}\vec{m}+\frac{1}{12}}}{(q)_{\vec{m}}}
\end{align}
\end{subequations}
and also
\be
\chi_{1,2} = \sum_{\vec{m}\vin{2}}\frac{q^{\frac{1}{2}\vec{m}^t\bsm 1 & 0\\0 & 1\esm\vec{m}+\frac{1}{2}\bsm 1\\1\esm^{\!{\!t}} \vec{m}}}{(q)_{\vec{m}}} \ .
\ee
Using the equality to the bosonic representation of the characters, these give \emph{bosonic-fermionic q-series identities} generalizing the left and right hand sides of \eqref{chars-bosonic}.
In \eqref{fe-h12} to \eqref{fe-h14}, also the last line of \eqref{fe-theta_by_eta} may be used, where $m_1+m_2 \equiv 1 \pmod{2}$.

It is remarkable that, although two of the characters have inhomogeneous modular weight, there is a uniform representation for all five characters with the same matrix $A$ in every case.
But on the other hand, this is a satisfying result, since this is also the case for all other models for which fermionic character expressions are known: Their different modules are only distinguished by the linear term in the exponent, not by the quadratic one. Note that the fact that the quadratic form is diagonal goes well with the description of the $c=-2$ model in terms of symplectic fermions \cite{Kau95,Kau00}, see section \ref{ref:quasi-w2333}.

The results are also in agreement with Nahm's conjecture (see section \ref{ref:nahm}), which predicts that for a matrix of the form $A=\bsm\alpha & 1-\alpha \\ 1-\alpha & \alpha\esm$ with rational coefficients, there exist a vector $\vec{b}\in\mathbb{Q}^r$ and a constant $c\in\mathbb{Q}$ such that $f_{A,\vec{b},c}(\tau)=\sum_{\vec{m}\in(\mathbb{Z}_{\geq 0})^r}^{\infty}\frac{q^{\frac{1}{2}\vec{m}^tA\vec{m}+\vec{b}^t\vec{m}+c}}{(q)_{\vec{m}}}$ is a modular function.

\subsection{Fermionic Character Expressions for $\mathcal{W}(2,2p-1,2p-1,2p-1)$}
\label{ref:general}

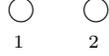
\begin{figure}\centering
\setlength{\unitlength}{1cm}
\begin{picture}(3.5,2)(-1,0)
\put(0,0.5){\circle{0.3}}
\put(1,0.5){\circle{0.3}}
\put(-0.1,0){$\scriptstyle 1$}
\put(0.9,0){$\scriptstyle 2$}
\end{picture}
\label{fig:d2}
\caption{The Dynkin diagram of $D_2=A_1\times A_1$}
\end{figure}

\begin{figure}\centering
\setlength{\unitlength}{1cm}
\begin{picture}(7,2)(-1,0)
\multiput(0,0.5)(1,0){3}{\circle{0.3}}
\multiput(0.3,0.5)(1,0){2}{\line(1,0){0.4}}
\put(2.25,0.4){$\cdots$}
\put(3,0.5){\circle{0.3}}
\put(4,1){\circle{0.3}}
\put(4,0){\circle{0.3}}
\put(3.3,0.5){\line(1,1){0.4}}
\put(3.3,0.5){\line(1,-1){0.4}}
\put(-0.1,0){$\scriptstyle 1$}
\put(0.9,0){$\scriptstyle 2$}
\put(1.9,0){$\scriptstyle 3$}
\put(2.7,0){$\scriptstyle n-2$}
\put(4.3,0.9){$\scriptstyle n-1$}
\put(4.3,-0.1){$\scriptstyle n$}
\end{picture}
\label{fig:d-generalize}
\caption{The Dynkin diagram of $D_n$}
\end{figure}
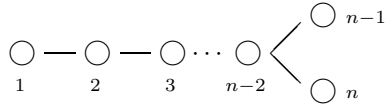

The matrix $\frac{1}{2}\bsm1 & 0 \\ 0 & 1 \esm$ was found in the quadratic form of the fermionic expressions for the $\mathcal{W}(2,3,3,3)$ model at $c=-2$ in the previous section. A generalization to $\mathcal{W}(2,2p-1,2p-1,2p-1)$ is possible by recognizing that the matrix in the case of $p=2$ is just the inverse of the Cartan matrix of the degenerate case $D_2=so(4)=A_1 \times A_1$ of the $D_n=so(2n)$ series of simple Lie algebras, where the corresponding Dynkin diagram consists just of two disconnected nodes, as shown in figure \ref{fig:d2}.
Consequently, one may try the inverse Cartan matrices
\be
C_{D_p^{-1}}=\begin{pmatrix}
1		& 1		& \cdots	& 1		& \frac{1}{2}		& \frac{1}{2}		\\
1		& 2		& \cdots	& 2		& 1			& 1			\\
\vdots		& \vdots	& \ddots	& \vdots	& \vdots		& \vdots		\\
1		& 2		& \cdots	& p-2		& \frac{p-2}{2}	& \frac{p-2}{2}	\\
\frac{1}{2}	& 1		& \cdots	& \frac{p-2}{2}& \frac{p}{4}		& \frac{p-2}{4}	\\
\frac{1}{2}	& 1		& \cdots	& \frac{p-2}{2}& \frac{p-2}{4}	& \frac{p}{4}
\end{pmatrix}
\ee
of $D_p=so(2p)$, $p>2$, for the fermionic expressions of the characters of the $c_{p,1}$ models in the case of $p>2$. The first thing we noticed by comparing expansions when we tried these matrices in \eqref{nahm} is that $\vec{b}=0$ leads to a fermionic expression for $\frac{\Theta_{0,p}}{\eta(q)}$. However, the restriction $m_1+m_2\equiv 0\pmod{2}$ has to be changed to $m_{p-1}+m_p\equiv 0\pmod{2}$ implying that particles of the two species corresponding to the two nodes labeled by $n-1$ and $n$ in the $D_n$ Dynkin diagram (see figure \ref{fig:d-generalize}),
which are both connected to the node labeled by $n-2$, may only be created in pairs, as will be shown in detail in section \ref{ref:quasi-fff}. These expressions coincide with the ones found in \cite{KKMM93a} (but only the ones with $\vec{b}=0$), since the characters of the free boson with central charge $c=1$ and compactification radius $r=\sqrt{\frac{p}{2}}$ \cite{Gin88} equal some of the characters of the $c_{p,1}$ models. The expressions for $\frac{\Theta_{\lambda,p}}{\eta(q)}$ have $+\frac{\lambda}{2}$ and $-\frac{\lambda}{2}$ in the last two entries of $\vec{b}$ and zero in the other components as is the case for the other, strictly two-dimensional fermionic expression for $\frac{\Theta_{\lambda,p}}{\eta(q)}$ given earlier in \eqref{fe-theta_by_eta}.

Still missing now are fermionic expressions for those characters whose bosonic form is of inhomogeneous modular weight, i.e.~which consist of theta and affine theta functions. For the vacuum character of $c=-2$ the vector is $\vec{b}=\frac{1}{2}\bsm 1 \\ 1 \esm$. Based on experience with fermionic expressions for other models, one may guess that the vector for the inhomogeneous characters of any $c_{p,1}$ model will have $+\frac{\lambda}{2}$ in both its last components, while the rest of the $k-2$ components will increase in integer steps from top to bottom, starting with zero at the component number $i$. The value of $i$ depends on the values of $\lambda$ and $k$.
All components above the component number $i$ are zero, too. The detailed description of this vector in dependence of $\lambda$ and $k$ is given below. In this way, expressions for all characters of all $c_{p,1}$ models can be found and thus a whole new, infinite set of bosonic-fermionic $q$-series identities, also given below. In section \ref{ref:quasi-w2333} and \ref{ref:quasi-triplet}, we will propose a physical interpretation in terms of quasi-particles.
Expanding the new fermionic character expressions in $q$, one may convince oneself that all coefficients match those of the bosonic character expressions. The identities reviewed here have been discovered by the present authors in collaboration with Carsten Grabow in \cite{FGK07}, and subsequently a proof has been published in \cite{War07}.

To summarize, the fermionic sum representations for all characters of the $\mathcal{W}(2,2p-1,2p-1,2p-1)$, $p\geq 2$, series of triplet algebras corresponding to central charge $c_{p,1}$ can be expressed as follows and indeed equal the bosonic ones (cf. \eqref{triplet-char-0}-\eqref{tilde_char_minus}), the latter being redisplayed on the right hand side for convenience \cite{FGK07}:
\begin{subequations}
\begin{alignat}{2}
\label{fe-1}\chi_{\lambda,k} & = \sum_{\substack{\vec{m}\vin{k} \\ m_{k-1}+m_k\equiv 0 \pmod{2}}}\mspace{-8mu} \frac{q^{\vec{m}^t  C_{D_{k}}^{-1}\vec{m}+\vec{b}_{\lambda,k}^t\vec{m}+c^{\star}_{\lambda,k}}}{(q)_{\vec{m}}} && = \frac{\Theta_{\lambda,k}}{\eta} \\
\label{fe-2}\chi_{\lambda',k}^{+} & =  \sum_{\substack{\vec{m}\vin{k} \\ m_{k-1}+m_k\equiv 0 \pmod{2}}}\mspace{-8mu}\frac{q^{\vec{m}^t C_{D_{k}}^{-1}\vec{m}+\vec{b'}_{\lambda',k}^{+^{t}}\vec{m}+c^{\star}_{\lambda',k}}}{(q)_{\vec{m}}} && =  \frac{(k-\lambda')\Theta_{\lambda',k}+(\partial\Theta)_{\lambda',k}}{k\eta} \\
\label{fe-3}\chi_{\lambda',k}^{-} & =  \sum_{\substack{\vec{m}\vin{k} \\ m_{k-1}+m_k\equiv 1 \pmod{2}}}\mspace{-8mu}\frac{q^{\vec{m}^t C_{D_{k}}^{-1}\vec{m}+\vec{b'}_{\lambda',k}^{-^t}\vec{m}+c^{\star}_{k-\lambda',k}}}{(q)_{\vec{m}}} && = \frac{\lambda'\Theta_{\lambda',k}-(\partial\Theta)_{\lambda',k}}{k\eta}
\end{alignat}\end{subequations}
for $0 \leq \lambda \leq k$ and $0 < \lambda' < k$, where $k=p$ since $p'=1$ and $(\vec{b}_{\lambda,k})_i=\frac{\lambda}{2}(\pm\delta_{i,k-1}\mp\delta_{i,k})$ for $1 \leq i \leq k$, $(\vec{b'}_{\lambda',k}^{+})_i=\max\{0,\lambda'-(k-i-1)\}$ for $1 \leq i < k-1$ and $(\vec{b'}_{\lambda',k}^{+})_i=\frac{\lambda'}{2}$ for $k-1 \leq i \leq k$, $(\vec{b'}_{\lambda',k}^{-})_i=(\vec{b'}_{k-\lambda',k}^{+})_i$ and $c^{\star}_{\lambda,k}=\frac{\lambda^2}{4k}-\frac{1}{24}$. Note that this means the characters and not the torus vacuum amplitudes \eqref{tilde_char_plus} and \eqref{tilde_char_minus}. Note that $\lim\limits_{\alpha\to 0}\tilde{\chi}_{\lambda,k}^+=\lim\limits_{\alpha\to 0}\tilde{\chi}_{\lambda,k}^-=\chi_{\lambda,k}$ for $0<\lambda<k$. Note also that in \eqref{fe-1}, also $m_{k-1}+m_k\equiv 1 \pmod{2}$ may be used as restriction, but then the vector and the constant change to $\vec{b}_{k-\lambda,k}$ and $c^{\star}_{k-\lambda,k}$, respectively (cf. \eqref{fe-theta_by_eta}).
Thus, as in the previous section, the $p\times p$ matrix $A=C_{D_p}^{-1}$ is the same for all characters corresponding to a fixed $p$, i.e.~for a fixed model. This is in agreement with previous results on fermionic expressions, since it is known to also be the case for the characters of a given minimal model (see e.g.~\cite{Wel05}).

For example, the fermionic expression of the vacuum character of the theory corresponding to central charge $c_{5,1}=-18.2$ would be
\be
\chi^+_{4,5}=\frac{\Theta_{4,5}+(\p\Theta)_{4,5}}{5\eta}=\sum_{\substack{\vec{m}\vin{5} \\ m_4+m_5 \equiv 0 \pmod{2}}}\frac{q^{\vec{m}^t\bsm 1 & 1 & 1 & \frac{1}{2} & \frac{1}{2} \\ 1 & 2 & 2 & 1 & 1 \\ 1 & 2 & 3 & \frac{3}{2} & \frac{3}{2} \\ \frac{1}{2} & 1 & \frac{3}{2} & \frac{5}{4} & \frac{3}{4} \\ \frac{1}{2} & 1 & \frac{3}{2} & \frac{3}{4} & \frac{5}{4} \esm \vec{m}+\bsm 1 \\ 2 \\ 3 \\ 2 \\ 2\esm^t\vec{m}+\frac{91}{120}}}{(q)_{\vec{m}}} \ .
\ee

\section{Quasi-Particle Interpretation of the Triplet $\mathcal{W}$-Algebras}
\label{ref:quasi-fff}

\subsection{Quasi-Particle Interpretation}

Non-unique realizations of the state spaces in two-dimensional conformal field theories establish the existence of several alternative character formulae. 

The original formula, the bosonic representation (cf.~section \ref{ref:chars-fermionic}), which traces back to Feigin and Fuks \cite{FF83} and Rocha-Caridi \cite{RoC84}, is directly based upon the structure of null vectors, i.e.~the invariant ideal is divided out. The occurrence of a factor $(q)_\infty$ in the denominator arises naturally in the construction of Fock spaces using bosonic generators. Indeed, the character of a free chiral boson is given by
\be\label{free-boson}
\chi_{B}=\sum_{n=0}^{\infty}p(n)q^n=\prod_{n=1}^{\infty}(1-q^n)^{-1}=\frac{1}{(q)_{\infty}} \ ,
\ee
where $p(n)$ is the number of additive partitions of the integer $n$ into integer parts greater than zero which don't have to be distinct. Encoded by the numerator, these spaces are then truncated in a particular way in the general bosonic character expression. The interpretation as partition functions requires these expressions to be modular covariant, which is easily checked when expressing the characters in terms of $\Theta$-functions (cf.~section \ref{ref:theta}).

In contrast, the fermionic representations possess a remarkable interpretation in terms of quasi-particles for the states, obeying Pauli's exclusion principle. The character of a free chiral fermion with periodic or anti-periodic boundary conditions is given respectively by
\begin{subequations}
\begin{align}
\label{free-fermion-p}\chi_{F,P} & =\sum_{n=0}^{\infty}\frac{q^{\frac{1}{2}m^2-\frac{1}{2}m}}{(q)_m}\quad\text{or} \\
\label{free-fermion-a}\chi_{F,A} & =\sum_{n=0}^{\infty}\frac{q^{\frac{1}{2}m^2}}{(q)_m} \ .
\end{align}
\end{subequations}
In the following section, it will be discussed how this comes about.

The bosonic representations are in general unique, since a natural level gradation in terms of the eigenvalue of $L_0$ is induced by the operators $L_n$. Although the fermionic ones are also obviously graded by their $L_0$ eigenvalue, there is in general more than one fermionic expression for the same character, since different types of generalized exclusion statistics may be imposed which might force different quasi-particle systems to lead to the same fermionic character expression.

\subsection{Quasi-Particle Representation of Fundamental Fermionic Forms}\label{ref:quasi-fff2}

The general fermionic character expression is a linear combination of fundamental fermionic forms. The characters of various series of rational CFTs, including the $ c_{p,1} $ series, can be represented as a single fundamental fermionic form \cite{Wel05,BMS98,DKMM94}. For simplicity, we won't deal with the most general case here, but with a certain specialization. This specialization is also called fundamental fermionic form in \cite{BMS98}.

Fermionic sum representations for characters admit an interpretation in terms of fermionic quasi-particles, as shown in \cite{KM93} (see also \cite{KKMM93a}). This can be easily seen from the fundamental fermionic form
\be \label{quasi-fff}
\chi(q)=\sum_{\substack{\vec{m}\vin{r} \\ \text{restrictions}}}q^{\vec{m}^tA\vec{m}+\vec{b}^t\vec{m}}
\prod_{a=1}^{r}\begin{bmatrix}((\mathds{1} -2A)\vec{m}+\vec{u})_a\\m_a\end{bmatrix}_q \ ,
\ee
with the help of combinatorics: The number of additive partitions $ P_M(N,N') $ of a positive integer $ N $ into $ M $ distinct non-negative integers which are smaller than or equal to $ N' $ is stated by \cite[p. 23]{Sta72}
\be \label{distinct}
\sum\limits_{N=0}^{\infty}P_M(N,N')q^N=q^{\frac{1}{2}M(M-1)}\begin{bmatrix}N'+1 \\ M \end{bmatrix}_q \ ,
\ee
which in the limit $ N' \to \infty $ takes the form
\be \label{distinct-parts}
\lim_{N'\to\infty}\sum\limits_{N=0}^{\infty}P_M(N,N')q^N=q^{\frac{1}{2}M(M-1)}\frac{1}{(q)_M} \ .
\ee
(A possible constant $c$ has been omitted, since it would just result in an overall shift of the energy spectrum of the resulting quasi-particles.) This formula is tailored to our needs, because the requirement of distinctiveness expresses the fermionic nature of the quasi-particles, i.e.~Pauli's exclusion principle.
To make use of \eqref{distinct-parts}, \eqref{quasi-fff} can be reformulated to
\begin{align}
\chi(q) & =\sum_{\substack{\vec{m}\vin{r} \\ \text{restrictions}}}^{\infty}q^{\frac{1}{2}\sum_{i=1}^r (m_i^2-m_i)+\sum_{i=1}^r (b_i+\frac{1}{2})m_i+\sum_{i,j=1}^r A_{ij}m_im_j-\frac{1}{2}\sum_{i=1}^r m_i^2}\nn\\
&\qquad\qquad\times 
\prod_{a=1}^{r}\begin{bmatrix}((\mathds{1} -2A)\vec{m}+\vec{u})_a\\m_a\end{bmatrix}_q \nonumber\\
& = \prod_{i=1}^r \left(\sum_{\substack{m_i\\ \text{restrictions}}}^{\infty}q^{\frac{1}{2}\sum_{i=1}^r (m_i^2-m_i)+(b_i+\frac{1}{2})m_i+\sum_{j=1}^r A_{ij}m_im_j-\frac{1}{2}m_i^2}\right) \nn\\
&\qquad\qquad\times\prod_{a=1}^{r}\begin{bmatrix}((\mathds{1} -2A)\vec{m}+\vec{u})_a\\m_a\end{bmatrix}_q \ .
\end{align}
Applying \eqref{distinct-parts} to the fundamental fermionic form \eqref{quasi-fff} leads to
\be \label{zwischen-general}
\prod_{i=1}^r \left(\sum_{\substack{m_i\\ \text{restrictions}}}^{\infty}\sum_{N=0}^{\infty}P_{m_i}(N,\left( (\mathds{1} -2A)\vec{m}+\vec{u}\right) _a -1)q^{N+(b_i+\frac{1}{2})m_i+\sum_{j=1}^r A_{ij}m_im_j-\frac{1}{2}m_i^2}\right) \ .
\ee
We can then make use of the relation
\be \label{distinct-parts-relation}
\sum_{N=0}^{\infty}P^0_M(N,N')q^{N+kM}=\sum_{N=0}^{\infty}P^k_M(N,N'+k)q^N \ ,
\ee
where we defined $P^k_M(N,N')$ like $P_M(N,N')$ but with the additional requirement that all the integers that make up a partition have to be greater than or equal to $k$. This relation is obvious since it is a one-to-one mapping of partitions and thus nothing more than just a mere shift of the partitions: Each part of a given partition of $N$ into $M$ distinct parts is increased by $k$, which turns $N$ into $N+Mk$.
\eqref{distinct-parts-relation} allows us to rewrite \eqref{zwischen-general} into
\be \label{zwischen-general2}
\prod_{i=1}^r \left(\sum_{\substack{m_i\\ \text{restrictions}}}^{\infty}\sum_{N=0}^{\infty}P^{b_i+\frac{1}{2}+ \left( (A-\frac{1}{2}\mathds{1})\vec{m}\right) _i}_{m_i}(N,-(A-\frac{1}{2}\mathds{1} ) \vec{m}) _a+\vec{b}_a-\frac{1}{2}+\vec{u}_a)q^N\right) \ .
\ee
For the quasi-particle interpretation, the characters are regarded as partition functions $Z$ for left-moving excitations with the ground-state energy scaled out
\be \label{partition_function}
\chi \sim Z=\sum\limits_{\text{states}}\e^{-\frac{E_{\text{states}}}{kT}}=\sum\limits_{l=0}^{\infty}P(E_l)\e^{-\frac{E_l}{kT}}
\ee
with $T$ being the temperature, $k$ the Boltzmann's constant, $E_l$ the energy and $P(E_l)$ the degeneracy of the particular energy level $l$.

The energy spectrum consists of all the excited state energies (minus the ground state energy) that are given by 
\be
\label{quasispec}
E_l=E_{ex}-E_{GS}=\sum_{i=1}^r\sum_{\substack{\alpha=1 \\ \text{restrictions}}}^{m_i}e_i(p^i_{\alpha}) \ ,
\ee
and the corresponding momenta of the states are given by
\be
P_ {ex}=\sum_{i=1}^r\sum\limits_{\substack{\alpha=1 \\ \text{restrictions}}}^{m_i}p_{\alpha}^i \ ,
\ee
where $r$ denotes the number of different species of particles, $m_i$ the number of particles of species $i$ in the state, $e_i(p_{\alpha}^i) $ the single-particle energy of the quasi-particle particle $\alpha$ of species $i$ and the subscript `restrictions' indicates possible rules under which the excitations may be combined.
\eqref{quasispec} is referred to as a quasi-particle spectrum in statistical mechanics (see e.g.~\cite{McC94}). Quasi means in this context that for example magnons or phonons have other properties than real particles like protons or electrons. And in addition, the spectrum above may contain single-particle energy levels that are different from the form in relativistic quantum field theory
$ e_\alpha(p)=\sqrt{M_\alpha^2+p^2} $.
This means that if we assume massless single-particle energies
\be \label{single-particle-energy}
e_i(p^i_{\alpha})=e(p^i_{\alpha})=vp^i_{\alpha}
\ee
($v$ referred to as the fermi velocity, spin-wave velocity, speed of sound or speed of light), where $p_{\alpha}^i$ denotes the quasi-particle $\alpha$ of ``species'' $i\ (1\leq i \leq r)$, and if in \eqref{zwischen-general2} we set 
\be \label{boltzmann}
q=\e^{-\frac{v}{kT}} \ ,
\ee
we can read off that the partition function corresponds to a system of quasi-particles that are of $r$ different species and which obey the Pauli exclusion principle
\be
p_{\alpha}^i \neq p_{\beta}^i  \quad \text{for} \ \alpha \neq \beta \quad \text{and all}\quad i \ ,
\ee
in order to satisfy Fermi statistics, but whose momenta $p_{\alpha}^i$ are otherwise freely chosen from the sets
\be
P_i=\left\lbrace  p_{\text{min}}^i, p_{\text{min}}^i+1,p_{\text{min}}^i+2,\ldots,p_{\text{max}}^i \right\rbrace
\ee
with minimum momenta
\be
\label{pmin}p_{\text{min}}^i(\vec m)=\left[((A-\frac{1}{2})\vec m)_{i}+b_{i}+\frac{1}{2}\right]
\ee
and with the maximum momenta
\be
\label{pmax}p_{\text{max}}^i(\vec m)=-((A-\frac{1}{2}\mathds{1})\vec{m})_i+(\vec{b})_i-\frac{1}{2}+(\vec{u})_i=-p_{\text{min}}^i(\vec m)+2(\vec{b})_i+(\vec{u})_i \ .
\ee
Thus, $p_{\text{max}}^i$ is either infinite if $(\vec{u})_i$ is infinite or finite and dependent on $\vec{m}$, $A$, $(\vec{b})_i$ and $(\vec{u})_i$. Note that if $(\vec{u})_i$ is infinite for all $i\in\{1,\ldots,r\}$, then \eqref{quasi-fff} reduces to the form \eqref{nahm} of Nahm's conjecture. Of course, \eqref{quasi-fff} is only an often encountered specialization of the most general fundamental fermionic form, since the components of the $q$-binomial coefficient may be of a different shape than that given in \eqref{quasi-fff}, but the generalization of the previous steps is obvious.
To sum up, this means that a multi-particle state with energy $E_l$ may consist of exactly those combinations of quasi-particles of arbitrary species $i$, whose single-particle energies $e(p^i)$ add up to $E_l$ and where Pauli's principle holds for any two quasi-particles of that combination which belong to the same species. Possible sum restrictions then result in the requirement that certain particles may only be created in conjunction with certain others. Thus, the characters \eqref{free-fermion-p} and \eqref{free-fermion-a} of the free chiral fermion with respectively periodic or anti-periodic boundary conditions are obtained in the case of $r=1$, $p_{\text{max}}=\infty$, $p_{\text{min}}^{(P)}=0$ and $p_{\text{min}}^{(A)}=\frac{1}{2}$. On the other hand, the character \eqref{free-boson} of a free chiral boson is obtained by setting $r=1$, $p_{\text{min}}=1$ and $p_{\text{max}}=\infty$ and simply not imposing any exclusion rules, i.e.~not using \eqref{distinct}.

Although the upper momentum boundaries may seem artificial, the phenomenon that the momenta $p^{\alpha}_{i_\alpha}$ for $2 \le \alpha \le n$ are restricted to take only a finite number of values for given $\vec{m}$ is a common occurrence in quantum spin chains.

\subsection{The $c=-2$ Model} \label{ref:quasi-w2333}

In section \ref{ref:general}, we reviewed the fermionic character expressions for the series of triplet $\mathcal{W}$-algebras \cite{FGK07}. In this and in the following section, we discuss the quasi-particle content, which one can derive from the fermionic expressions, of the $c_{p,1}$ logarithmic conformal field theories, which have $\mathcal{W}(2,2p-1,2p-1,2p-1)$ as symmetry algebras.

We start with the case $ p=2 $, i.e.~$ c_{2,1}=-2 $. In contrast to the characters for e.g.~the minimal models, these characters are the traces over the representation modules of the triplet $\mathcal{W}$-algebra, instead of the Virasoro algebra only. However, although highest weight states are labeled by two highest weights in this case, $h$ and $w$ as the eigenvalues of $L_0$ and $W_0$ respectively, we consider only the traces of the operator $q^{L_0-\frac{c}{24}}$. It turns out that these $\mathcal{W}$-characters are given as infinite sums of Virasoro characters, for example \cite{Flo96}
\be
\chi_{\ket{0}} = \sum_{k=0}^\infty (2k+1)\chi_{\ket{h_{2k+1,1}}}^{\text{Vir}} \ .
\ee

Let us now come to the vacuum character \eqref{fe-h11} for the $c_{2,1}$ model, which features the interesting sum restriction $ m_1+m_2\equiv 0 \pmod{2} $ expressing the fact that particles of type $1$ and $2$ must be created in pairs. Thus, the existence of one-particle states for either particle species is prohibited. Therefore, the single-particle energies must be extracted out of the observed multi-particle energy levels.

Applying \eqref{distinct-parts} to the fermionic sum representation (cf. also \eqref{fe-h11})
\be
q^{-\frac{1}{12}}\chi_{1,2}^{+} = \sum_{\substack{\vec{m}\vin{2} \\ m_1+m_2\equiv 0 \pmod{2}}}\frac{q^{\frac{1}{2}\vec{m}^t\bsm 1 & 0\\0 & 1 \esm\vec{m}+\frac{1}{2}\bsm 1\\ 1 \esm^{\!{\!t}}\vec{m}}}{(q)_{\vec{m}}}
\ee
of the vacuum character leads to
\be \label{zwischen}
\chi_{1,2}^+=\biggl( \sum_{m_1=0}^{\infty}\sum_{N=0}^{\infty}P_{m_1}(N)q^{N+m_1}\biggr) \biggl( \sum_{\substack{m_2=0 \\ m_2 \equiv m_1 \pmod{2}}}^{\infty}\sum_{N=0}^{\infty}P_{m_2}(N)q^{N+m_2}\biggr) \ ,
\ee
where the constant $c$ has been omitted, since it would just result in an overall shift of the energy spectra. Using massless single-particle energies \eqref{single-particle-energy} and setting \eqref{boltzmann} in \eqref{zwischen-general2} then results in the partition function \eqref{partition_function} corresponding to a system of two quasi-particle species, with both species having the momentum spectrum $\N_{\geq 1}$, i.e.~a multi-particle state with energy $E_l$ may consist of exactly those combinations of an even number of quasi-particles, having momenta $p_{\alpha}^i\ (i\in \left\lbrace 1,2 \right\rbrace )$, whose single-particle energies $e(p_{\alpha}^i)$ add up to $E_l$ and where the momenta $p_{\alpha}^i\in\N_{\geq 1}$ of each two of the quasi-particles in that combination are distinct unless they belong to different species, i.e.~they respect the exclusion principle. Formally, these spectra belong to two free chiral fermions with periodic boundary conditions. Note in this context the physical 
interpretations in \cite{Kau95,Kau00}, in which the CFT for $ c_{2,1}=-2 $ is generated from a symplectic fermion, a free two-component fermion field of spin one.

\subsection{The $p > 2$ Relatives} \label{ref:quasi-triplet}

Besides the best understood LCFT with central charge $c_{2,1}=-2$, we now have
a look at the quasi-particle content of its $c_{p,1}$ relatives.

The restrictions $m_{p-1}+m_{p}\equiv Q \pmod{2}$ ($Q$ can be thought of as denoting the total charge of the system) in \eqref{fe-1} to \eqref{fe-3} imply that the quasi-particles of species $p-1$ and $p$ are charged under a $\Z_2$ subgroup of the full symmetry of the $D_p$ Dynkin diagram \cite{KKMM93a}, while all the others are neutral. This charge reflects the $\mathfrak{su}(2)$ structure carried by the triplet $\mathcal{W}$-algebra such that all representations must have ground states, which are either $\mathfrak{su}(2)$ singlets or $\mathfrak{su}(2)$ doublets. In comparison to the $c_{2,1}=-2$ model, there exist $p$ quasi-particles in each member of the $c_{p,1}$ series, exactly two of which can only be created in pairs, while the others do not have this restriction. 
These observations suggest the following conjecture: The $c_{p,1}$ theories might possess a realization in terms of free fermions such that they are generated by one pair of symplectic fermions and $p-2$ ordinary fermions. Realizations of that kind are unknown so far, except for the well-understood case $p=2$, and might constitute an interesting direction of future research.

Contrary to the $p=2$ case, the quasi-particles do not decouple here: The minimal momenta for the quasi-particle species can be read off \eqref{pmin} and are given by
\be
p_{\text{min}}^i(\vec m)=\begin{cases}-\frac{1}{2}m_i+\sum_{j=1}^{i}jm_j+\sum_{j=i+1}^{p-2}im_j+\frac{i}{2}(m_{p-1}+m_p)+i+\frac{1}{2} \text{ for } 1\leq i \leq p-2 \\
-\frac{1}{2}m_i+\sum_{j=1}^{p-2}\frac{j}{2}m_j+\frac{p-1}{2}+\frac{1}{2}+\begin{cases}(\frac{p}{4}m_{p-1}+\frac{2-p}{4}m_p) \text{ for } i=p-1 \\
(\frac{2-p}{4}m_{p-1}+\frac{p}{4}m_p) \text{ for } i=p \end{cases}
\end{cases} \ .
\ee
Hence, they depend on the numbers of quasi-particles of the different species in the state. But as in the $p=2$ case, the momentum spectra are not bounded from above.

\section{Summary and Outlook}
\addcontentsline{toc}{chapter}{Summary and Outlook}

We mainly focused on the characters of the irreducible subrepresentations contained in the indecomposable Jordan cell representations of LCFTs.
Our aim was to highlight the crucial role of these particular characters for
our understanding of LCFT and for how LCFT fits into the larger landscape of (rational) CFTs. 

Firstly, the characters of the irreducible subrepresentations of indecomposable
representations are the only characters in a LCFT, which allow by studying their properties under modular transformations to deduce that the corresponding CFT is in fact logarithmic, and not an ordinary one. In fact, the character of the
vacuum representation of a LCFT generally is of this type.

Secondly, the modular differential equation can be constructed from the input of these particular characters alone. Its solutions yield a complete, finite dimensional representation of the modular group including the characters of the
irreducible subrepresentations of indecomposable representations and the necessary generalized character functions (which depend on both, $q$ and $\log(q)$, formally). The latter have an interpretation as torus vacuum amplitudes. 
The set of torus amplitudes yields, by a Verlinde-like formula, fusion rules in a limit, where the generalized character functions are projected out at the end. Equivalence of these fusion rules with rules computed by direct means has been established in \cite{FK07}.

Thirdly, in common LCFTs, the sector of the theory with $h=0$ is indecomposable with respect to the action of $L_0$. Thus the character of the irreducible subrepresentation of the indecomposable $h=0$ sector is an important tool to understand the vacuum structure of the corresponding LCFT. For example, the character for the vacuum representation is needed to prove the $C_2$ cofiniteness of the triplet series in the general approach taken in \cite{CF06}. The vacuum character will also play a key role in the study of
logarithmically extended minimal models. We showed that the modular differential equation predicts that augmented minimal models with central charge $c_{p,q}$ are only possible for conformal grids of size $(\alpha p-1)\times(\alpha q-1)$ with $\alpha$ odd. Furthermore, we computed the modular differential equation and its space of solutions for $c=c_{6,9}=0$, yielding a conjecture for the vacuum character of this theory, which still is only known to small finite order.

Finally, the characters of irreducible subrepresentations of indecomposable representations are precisely the characters, with which the $ADET$ classification and the Nahm conjecture extend in a natural way to LCFTs, when their fermionic sum representations are considered. Actually, this has been proven only for the $c_{p,1}$ series, but it presumably holds more generally for logarithmically extended minimal models.

We reviewed the derivation of fermionic expressions for all characters of each $c_{p,1}$ model \cite{FGK07}, the existence of which provides (in line with Nahm's conjecture) further evidence for the well-definedness of the logarithmic conformal field theories corresponding to central charge $c_{p,1}$ as being rational conformal field theories. Furthermore, we explained how these models admit an interpretation in terms of $p$ fermionic quasi-particle species. We reviewed how the obtained character expressions also lead to an infinite set of new bosonic-fermionic $q$-series identities generalizing \eqref{chars-bosonic}, see section \ref{ref:general}. The case at hand is special due to the inhomogeneous structure of the bosonic character expressions in terms of modular forms. Notwithstanding, there exist fermionic quasi-particle sum representations with the same matrix $A$ (cf. \eqref{nahm}) for each of the characters of each $c_{p,1}$ model. Namely, the matrix $A$ turns out to be the inverse of the Cartan 
matrix of the simply-laced Lie algebra $D_p$. Therefore, those expressions fit well into the known scheme of fermionic character expressions for other (standard) conformal field theories.

There are also numerous other avenues towards finding fermionic expressions. Among these are Bailey's lemma \cite{Bai49}, thermodynamic Bethe ansatz \cite{KNS93}, Kostka Polynomials and Hall-Littlewood Functions \cite{SW99b,DLT94,AKS05,War02,Mac79,Mac97}. A specific new technique involves quantum groups, crystal bases and finite paths \cite{HKKOTY98,HKOTY99,HKOTT01,Dek06}. The \emph{quantum groups}\index{quantum groups} \cite{Kas90,Kas95,HK02,SKAO96} deepen the understanding of symmetry in systems with an infinite number of degrees of freedom. In general, the connection of quantum groups to conformal field theory is given by the \emph{Kazhdan-Lusztig correspondence}\index{Kazhdan-Lusztig correspondence}. In particular, the Kazhdan-Lusztig-dual quantum group to the logarithmic $\mathcal{W}$-algebras is known (see e.g.~\cite{FGST05,FGST06a,FGST06b,FGST06c}).

Fermionic character expressions imply a realization of the underlying theory in terms of systems of fermionic quasi-particles. We detailed this for the case of the $\mathcal{W}(2,2p-1,2p-1,2p-1)$ series of triplet algebras. In the $c_{2,1}=-2$ model, i.e.~$p=2$, the quasi-particle interpretation implies that there exist two fermionic quasi-particle species whose members may only be created in pairs, i.e.~either a pair of particles from the same species or one particle from each species. This coincides with the realization of the $c=-2$ theory in terms of a pair of symplectic fermions \cite{Kau95,Kau00}, which is a free two-component fermion field of spin one. In the general $c_{p,1}$ model, there is a set of $p-2$ fermionic quasi-particle species, the members of which may -- aside from Pauli's exclusion principle -- be combined freely
in building an arbitrary multi-particle state, and additionally a set of two species, the members of which may only be created in pairs. This interpretation suggests that the $c_{p,1}$ theories might possess a realization in terms of free fermions such that they are generated by $p-2$ ordinary fermions and one pair of symplectic fermions. Such realizations had been unknown, except for $p=2$, and constitute a possible direction for further research.

\ack
The authors wish to thank for helpful discussions
Hendrik Adorf, 
Nils Carqueville,
Matthias Gaberdiel,
Carsten Grabow,
Muxin Han, 
Song He,
James Lepowsky,
Kirsten Vogeler, 
Ole Warnaar,
Trevor Welsh,
Robert Wimmer and
Martin Wolf. 
The authors are deeply grateful
to David Ridout for providing them the opportunity to contribute to this special
volume of J Phys A, and for helping so much getting deadline extension over
deadline extension. The anonymous referees deserve special thanks for their
very careful reading and more than helpful comments to considerably improve this
review. Finally, MK gratefully acknowledges the support of the European 
Research Council via the Starting Grant numbered 256994.

\vspace{3em}
\noindent {\bf References}
\vspace{1em}
\bibliographystyle{koehn-numbers}
\bibliography{master}
\end{document}